\documentclass[useAMS,usenatbib]{mn2e}

\usepackage{psfig}
\usepackage{natbib}
\usepackage{graphicx}
\usepackage{amsmath}
\usepackage{eepic}
\usepackage{epic}
\usepackage{picture}
\usepackage{amssymb}
\usepackage{multirow}
\title[Astronomical spectroscopy with EMCCDs.]{On the use of
  electron-multiplying CCDs for astronomical spectroscopy}

\author[S. M. Tulloch and V. S. Dhillon]{S. M. Tulloch$^{1}$\thanks{E-mail:
smt@qucam.com (SMT); vik.dhillon@sheffield.ac.uk (VSD)} and V. S.
Dhillon$^{1}$\footnotemark[1]\\
$^{1}$ Department of Physics and Astronomy,University of Sheffield, Sheffield S3 7RH, UK}

\begin{document}

\date{Accepted . Received 2010 ; in original form 2010}

\pagerange{\pageref{firstpage}--\pageref{lastpage}} \pubyear{2010}

\maketitle

\label{firstpage}

\begin{abstract}
Conventional CCD detectors have two major disadvantages: they are slow
to read out and they suffer from read noise. These problems combine to
make high-speed spectroscopy of faint targets the most demanding of
astronomical observations. It is possible to overcome these weaknesses
by using electron-multiplying CCDs (EMCCDs). EMCCDs are conventional
frame-transfer CCDs, but with an extended serial register containing
high-voltage electrodes. An avalanche of secondary electrons is
produced as the photon-generated electrons are clocked through this
register, resulting in signal amplification that renders the read
noise negligible. Using a combination of laboratory measurements with
the QUCAM2 EMCCD camera and Monte Carlo modelling, we show that it is
possible to significantly increase the signal-to-noise ratio of an observation by using an EMCCD, but only if it is optimised and utilised
correctly. We also show that even greater gains are possible through
the use of photon counting. We present a recipe for astronomers to
follow when setting up a typical EMCCD observation which ensures that
maximum signal-to-noise ratio is obtained. We also discuss the
benefits that EMCCDs would bring if used with the next generation of
extremely large telescopes. Although we mainly consider the spectroscopic use of EMCCDs, our conclusions are equally applicable to imaging.
\end{abstract}

%http://www.noao.edu/apj/keywords96.html
%%%%%%%%%%%%%%%%%%%%%%%%%%%%%%%%%%%%%%%%%%%%%%
\section{Introduction}

In 2001 a new type of detector was announced: the electron multiplying
CCD or EMCCD. This was first described by \cite{jerram01} at E2V
Technologies and \cite{Hynecek01} at Texas Instruments. EMCCDs
incorporate an avalanche gain mechanism that renders the electronic
noise in their readout amplifiers (known as {\em read noise})
negligible and permits the detection of single photon-generated
electrons (or {\em photo-electrons}). Whilst photon counting in the
optical has been possible for some time with image tube detectors,
such as the IPCS (\citealt{IPCS72}, \citealt{IPCS87}), and with
avalanche photodiode-based instruments, such as Optima
(\citealt{Kanbach2008}), it has never been available with the high
quantum efficiency (QE), large format and convenience of use of a CCD.

EMCCDs have generated a lot of interest in the high spatial-resolution
community (e.g. \citealt{Tubbs2002}), but
have received much less attention for other astronomical applications.
In this paper, we explore how EMCCDs can be best exploited for
spectroscopy, which is arguably the most important and fundamental
tool of astronomical research. Spectroscopy provides much more
information than photometry, such as the detailed kinematics, chemical
abundances and physical conditions of astronomical sources.  However,
compared to photometry, where the light from a star or distant galaxy
is concentrated onto a small region of the detector, spectroscopy
spreads the light from the source across the entire length of the
detector. The amount of light falling onto each pixel of the detector
is therefore much lower in spectroscopy than in photometry, which
makes the reduction of read noise much more important and implies that
EMCCDs should be ideally suited to this application.

Recognising the potential advantages of EMCCDs for astronomical
spectroscopy, two separate teams have recently constructed cameras for
this purpose and operated them at major observatories: QUCAM2 on the
ISIS spectrograph of the 4.2-m William Herschel Telescope (WHT; \citealt{Tulloch09}) and
ULTRASPEC on the EFOSC2 spectrograph of the 3.5-m New Technology
Telescope (\citealt{IvesUSPEC}, \citealt{2008AIPC..984..132D})
\footnote{See also \cite{2004SPIE.5491..677B}}. These detectors are
significantly harder to optimise and operate than conventional
CCDs due to the use of higher frame rates and the greater visibility of subtle
noise sources that can be overlooked in a conventional CCD.
In fact, incorrect operation can actually result in a worse signal-to-noise ratio (SNR) due to the
presence of \textit{multiplication noise} in the EMCCD (see Section \ref{sec:MNoise}). To date,
there has been no in-depth analysis of the performance of EMCCDs for
astronomical spectroscopy presented in the refereed astronomical
literature, and no guidance given to astronomers on how best to set up
an EMCCD observation in order to obtain maximum SNR. In Section 2, we give a brief
review of key EMCCD concepts. Section 3 describes the three observing
regimes that must be considered when using an EMCCD: conventional
mode, linear mode and photon-counting (PC) mode. The latter mode does not
lend itself to an analytic treatment, so in Section 4 we present the
results of Monte-Carlo modelling of the photon-counting performance of
EMCCDs. In Section 5, we provide a  recipe for obtaining maximum
SNR with an EMCCD camera. Finally, in Section 6 we look to the future
and demonstrate the advantages offered by EMCCDs when used for
spectroscopy on the proposed 42-m European Extremely Large Telescope
(E-ELT).

%%%%%%%%%%%%%%%%%%%%%%%%%%%%%%%%%%%%%%%%%%%%%%
\section{Key EMCCD concepts}

Although the principle of operation of an EMCCD is very similar to
that of a conventional CCD, there are some additional features that
need to be considered.

\subsection{EMCCD structure}
\label{sec:geom}

The structure of an EMCCD (Figure~1) has already been
described in some depth by \cite{Mackay}, \cite{Tulloch2004},
\cite{MarshHTRA} and \cite{IvesUSPEC}. Photo-electrons are transferred
into a conventional CCD serial register, but before reaching the
output amplifier they pass through an additional multi-stage register
(known as the \textit{electron-multiplication} or \textit{EM register}) where a
high-voltage (HV) clock of $>40$V produces a multiplication of the
photo-electrons through a process known as impact ionisation -- see
Figure~2. The EM output amplifier is similar to
that found in a conventional  CCD but is generally faster
and hence suffers from increased read noise.  Nevertheless a single
photo-electron entering the EM register will be amplified to such an
extent that the read noise is rendered insignificant and single
photons become clearly visible. Most EMCCDs also contain a
conventional low-noise secondary amplifier at the opposite end of
the serial register; use of this output transforms the EMCCD into a
normal CCD.

Most EMCCDs are of frame-transfer design -- see
Figure~1. Here, half the chip is covered with an
opaque light shield that defines a storage area. The charge in this
storage area can be transferred independently of that in the image
area. This allows an image in the storage area to be read out
concurrently with the integration of the next image, with just a few tens of millisecond
dead time between exposures. Incorporating frame-transfer architecture
into any CCD will greatly improve observing efficiency in high
frame-rate applications where the readout time is comparable to the
required temporal resolution (\citealt{DhillonUCAM07}). In the case of an EMCCD the use of
frame-transfer architecture is essential otherwise the SNR gains will
be nullified by dead time.

\begin{figure}
\begin{center}
\includegraphics[width=0.45\textwidth]{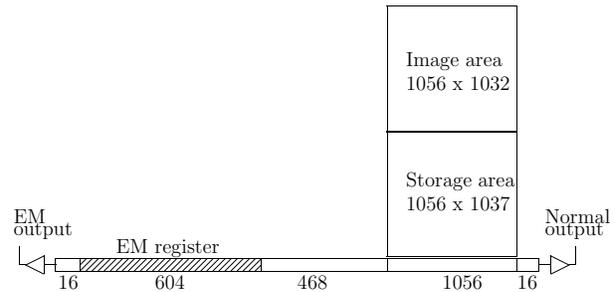}
\parbox{0.45\textwidth}{
\caption[Pipeline]{Schematic structure of an EMCCD, the E2V CCD201.
  Photo-electrons produced in the image area are vertically clocked
  downwards, first into the storage area, and then into the 1056-pixel
  serial register. For EMCCD operation, the charge is then
  horizontally clocked leftwards, through the 468-pixel extended serial
  register  and into the 604-pixel EM register, before being
  measured and digitised at the EM output. For conventional CCD
  operation, the charge in the serial register is horizontally clocked
  rightwards to the normal output.}}
\label{fig:pipeline}
\end{center}
\end{figure}

\begin{figure}
\begin{center}
\includegraphics[width=0.45\textwidth]{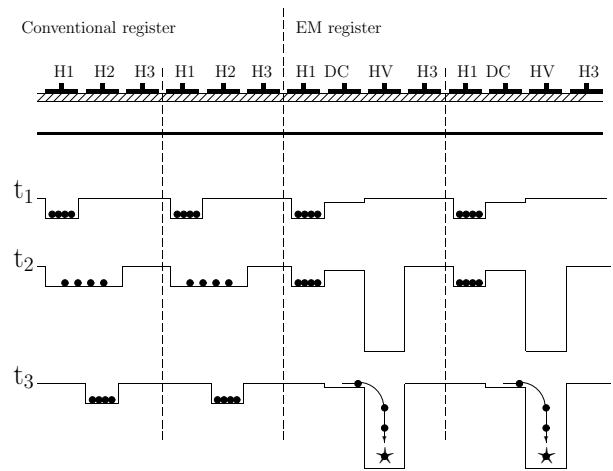}
\parbox{0.45\textwidth}{
\caption[Multiplication process]{The geometry of the serial and EM
  registers, showing how multiplication occurs. The top part of the
  diagram shows a cross-section through the EMCCD structure with the
  electrode phases lying at the surface. Below this are three
  snapshots showing the potential wells and the charge packets they
  contain at key moments (t$_1$,t$_2$,t$_3$) in the clocking
  process. At t$_3$ the photo-electrons undergo avalanche
  multiplication as they fall into the potential well below the HV
  clock phase. Note that this diagram does not show a complete pixel
  cycle.}}
\label{fig:electrodes}
\end{center}
\end{figure}

\subsection{Multiplication noise}
\label{sec:MNoise}
A single photo-electron entering the EM register can give rise to a
wide range of output signals. This statistical spread constitutes an
additional noise source termed multiplication noise
(\citealt{Hollenhorst}).  \cite{PCStrategies} derive the following
equation describing the probability $p(x)$ of an output $x$ from the
EM register in response to an input of $n$ (an integer) photo-electrons:
\begin{equation}
p(x)=\frac{x^{n-1}\exp(-x/g_A)}{g_A^n(n-1)!}{  }.
\label{eq:BasdenEquation}
\end{equation}
This is evaluated for several values of $n$ and with $g_A$ the EM gain (see Section 2.3) equal to 100, in
Figure~3,
\begin{figure}
\begin{center}
\includegraphics[width=0.45\textwidth]{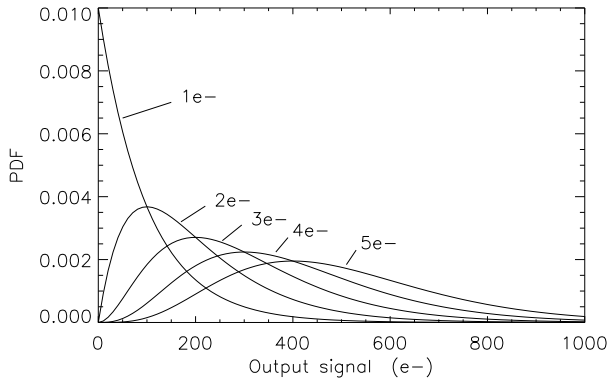}
\parbox{0.4\textwidth}{
\caption[Output of EM register]{Output of an EM register with $g_A=100$
  in response to a range of inputs from 1 to 5e$^-$. The $y$-axis
  shows the probability density function (PDF) of the output signal,
  i.e. the fraction of pixels lying within a histogram bin.}}
\label{fig:Electron12345e}
\end{center}
\end{figure}
which shows that for an output signal of 300 e$^-$, the input signal could
have been either 3 or 4e$^-$ with almost equal probability. The
overall effect is to double the variance of the signal, which is
statistically equivalent to halving the QE of the camera (see
Section~3.2). In the photon-noise dominated
regime this means that conventional CCDs will actually give a higher
performance. It is in the read-noise dominated regime that EMCCDs come
into their own, where their lack of read noise more than compensates
for the effects of multiplication noise. Note also that for signal
levels where there is a low probability of a pixel containing more
than one photo-electron it is possible to use a photon-counting
analysis of the image to remove the effect of multiplication noise
(see Section~3.3).

\subsection{Gain}
Astronomers typically refer to the gain (or strictly speaking system
gain, $g_S$) of a CCD camera as the number of photo-electrons
represented by 1 analogue-to-digital unit (ADU) in the raw image,
i.e. it has units of e$^-$/ADU. An EMCCD camera has another gain
parameter that we need to describe: the avalanche multiplication gain
$g_A$ (hereafter referred to as the \textit{EM gain}), and there is a risk of
confusion here with $g_S$. EM gain is simply a unitless multiplication
factor equal to the mean number of electrons that exit the EM register
in response to a single electron input. It is hence related to $g_S$
by the relation $g_A=g_{S0}/g_S$, where $g_{S0}$ is the system gain
(in units of e$^-$/ADU) measured with the EM gain set to unity.

\begin{figure}
\begin{center}
\includegraphics[width=0.5\textwidth]{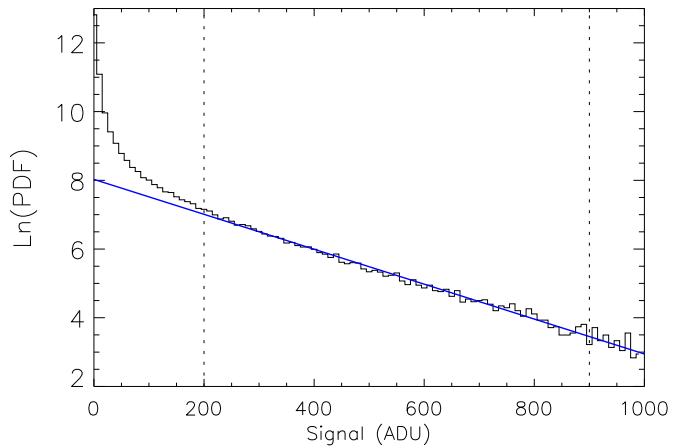}
\caption[EMCCD Gain calculation]{The histogram, plotted on a log$_e$
  vertical scale, of pixels in a weakly-illuminated EMCCD image. The
  solid line is a least-squares linear fit to the photo-electron events lying
  between the two vertical dotted lines. The slope of this fitted line
  can be used to calculate the system gain $g_S$ of the camera in e$^-$/ADU.}
% generated with "CalcGainEMCCD"
\label{fig:gaincalc}
\end{center}
\end{figure}

To measure the various EMCCD gain parameters, we need to first turn
off the EM gain by reducing the HV clock amplitude to 20V. At this
level the EM register will then behave as a conventional serial
register, i.e. 1 electron in, 1 electron out. We can now measure
$g_{S0}$, just as we would with a conventional CCD (there are various
methods, for example the photon transfer curve,
\citealt{Janesick}). To measure $g_S$, we weakly
($<$\,0.1e$^-$pix$^{-1}$) illuminate an EMCCD with a flat field so as to
avoid a significant number of pixels containing more than a single
electron. A histogram of such an image, with a vertical log scale, is
shown in Figure~\ref{fig:gaincalc}. In this histogram, the pixels
containing photo-electrons lie along a curve that is linear except at
low values where the effects of read noise become
dominant. Figure~\ref{fig:gaincalc} also shows a least-squares
straight-line fit to the linear part of the histogram: the gradient
of the line is equal to $-g_S$ (\citealt{Tulloch2004}). In this
particular case the camera had a system gain $g_S=0.005$ e$^-$/ADU,
i.e. a single photo-electron entering the EM register would produce a
mean signal of 200 ADU in the output image. When discussing noise
levels in an EMCCD it is more convenient to express this in units of
input-referenced photo-electrons (e$^-_{pe}$). So in the above
example, if the read noise is 5 ADU this would be quoted as
$5\times0.005= 0.025$e$^-_{pe}$.

\subsection{Clock-induced charge}
Clock-induced charge (CIC) is an important source of noise in
EMCCDs. Its contribution needs to be minimised.
 It consists of internally-generated electrons produced by
clock transitions during the readout process. CIC is visible in EMCCD
bias frames as a scattering of single electron events which at first
sight are indistinguishable from photo-electrons. It is only when a
histogram is made of the image that they appear different.

CIC is dependent on a number of factors. The amplitude of the clock
swings is relevant, as is the temperature (\citealt{Janesick}). At
first sight, one might assume that, since CIC is proportional to the
total number of clock transitions a pixel experiences during the
readout process, a pixel lying far from the readout amplifier should
experience a higher level of CIC. One would then expect CIC gradients
in both the horizontal and vertical axes of the image. This would be
true if the CCD is entirely cleared of charge prior to each readout.
However, this is never the case. One must consider that prior to each
readout the chip has either been flushed in a clear operation or read
out in a previous exposure. These operations leave a \textquoteleft
history\textquoteright{} of CIC events in the CCD pixels prior to our
subsequent measurement readout. The distribution of these historical
events will be higher the closer we get to the output amplifier
since the CIC charge residing in these pixels will have accumulated
through a larger number of clock transitions than for pixels more
distant from the amplifier. When these historical events are added to
the events created in the most recent measurement readout the overall effect is
that each pixel of the image will have experienced the same number of
clock transitions regardless of its position, and the resulting CIC
distribution will be flat.

\begin{figure}
\begin{center}
\includegraphics[width=0.5\textwidth]{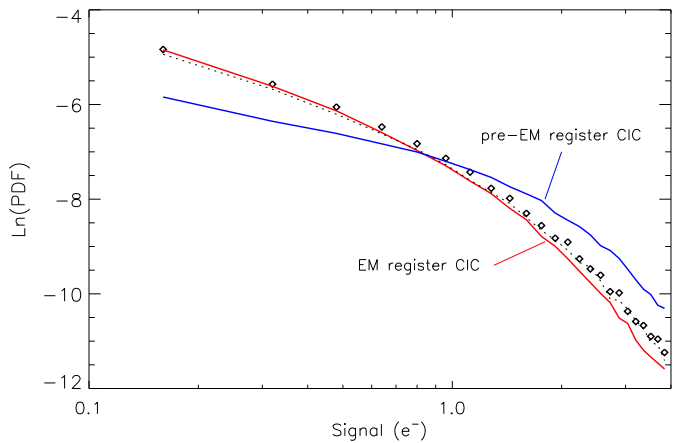}
%generated by "GenerateHistFitGraph"
\caption[Histogram QUCAM2]{Histogram (diamonds) of a QUCAM2 bias image
  compared with two models: pure in-register CIC and pure pre-EM
  register CIC. A fit to the data, consisting of 85\% CIC generated
  within the EM register and 15\% generated prior to the EM register
  is shown as a dotted curve.}
\label{fig:QUCAM2hist}
\end{center}
\end{figure}

CIC is
produced by both vertical and horizontal clocks, as well as the clocks
within the EM register. Vertical CIC can be virtually eliminated through the use of
non-inverted mode clocking (where the clock phases never fall more
than about 7 volts below the substrate of the CCD, the exact value depending on the CCD type). Serial-clock
CIC can be reduced by using lower clock amplitudes. CIC generated
within the EM register (described from here on as CICIR) is
harder to remove since any changes to the EM clock amplitudes produces
large changes in EM gain. A well optimised EMCCD will have its performance limited only by CICIR. Other noise sources such as amplifier read noise, dark current, image-area and serial-register CIC should all have been reduced to an insignificant level with respect to CICIR. This optimisation process for QUCAM2 is described in \cite{TullochPhD}.
As an example of this, the histogram of a bias image from
the QUCAM2 EMCCD camera is compared with the histograms of two other
images generated using a Monte Carlo model (see Section~4)
in Figure~\ref{fig:QUCAM2hist}. The
first of these models consists of CIC  originating prior to the
EM register, the second consists of CIC  originating at random
positions within the EM register. The latter will on average
experience less multiplication than the former since it will pass
through fewer stages of the multiplication register, producing a
histogram that shows an excess of low value pixels (as shown in
Figure~\ref{fig:QUCAM2hist}). Various other models were created with
mixes of the two noise sources. The best fit was found to correspond
to 85\% CICIR and 15\% pre-EM-register CIC, demonstrating that
QUCAM2 has been well optimised (at least as far as its CIC performance is
concerned with some other parameters such as CTE remaining non-optimal.)
and is dominated by CICIR. Details on the final performance of QUCAM2 and the values of some of its more critical parameters can be found in Table \ref{tab:QUCAM2_table}.
\begin{table}
\caption{QUCAM2 technical details.}
\begin{center}
    \begin{tabular}{  l  l  }
    \hline
    CCD type & E2V CCD201-20  \\
    Controller & ARC Gen. III  \\
    Operating temperature & 178K\\
    Pixel time (EM amplifier) & 1.3$\mu$s  \\
    Pixel time (normal amplifier) & 5.1$\mu$s  \\
    Frame transfer time & 13ms \\
    Row transfer time  & 12$\mu$s  \\
%    Full frame read (EM amplifier)  & 1.6s  \\
%    Full frame read (normal amplifier)  & 5.4s  \\
%    \hline
    EM multiplication gain $g_A$ & 1840\\
    HV clock rise-time & 70ns  \\
    HV clock fall-time & 150ns  \\
    HV clock voltage high & 40V (square wave)  \\
    Parallel clock voltages & -1/+8V\\
    Serial clock voltages & 0/+8.5V\\
    Substrate voltage & +4.5V \\
    Read-noise (EM amplifier)  & 40e$^-$ \\
    Read-noise (normal amplifier)  & 3.1e$^-$ \\
%    \hline
    Mean  charge in bias image & 0.013e$^-$pixel$^{-1}$ \\
    Cosmic ray rate & 0.9e$^-$pixel$^{-1}$hour$^{-1}$ \\
    Image area dark current & 1.5e$^-$pixel$^{-1}$hour$^{-1}$ \\
    EM register CIC probability & 1.4$\times$10e$^{-4}${ }transfer$^{-1}$ \\
    EM register (1e$^-$ level) CTE  & 0.99985 \\
    \hline
    \end{tabular}
\end{center}
\label{tab:QUCAM2_table}
\end{table}
Pre-EM register CIC electrons should not be confused with dark current
generated during the readout. These can easily outnumber CIC events if
the operational temperature is too high, if the CCD controller has
recently been powered on or if the CCD has recently been saturated to
beyond full-well capacity. The recovery time for these last two cases
is approximately 2 hours, and an accurate measurement of
CIC should not be attempted until after such a period.

CIC electrons generated within the EM register do not contribute as
much charge to an image as a CIC electron generated prior to the
register: CICIR has a fractional charge when expressed in units of
input-referenced photo-electrons. If we
assume that CICIR is generated randomly throughout the EM
register then we can calculate the average charge $\nu_C$ that it will
contribute to an image pixel. The calculation, derived in
Appendix~\ref{sec:fractionalApp}, shows that:
\begin{equation}
\nu_C\approx\frac{B_C}{\ln(g_A)}{  },
\end{equation}
where $B_C$ is the mean number of CICIR
events experienced by a pixel during its transit through the EM
register.
\subsection{Non-inverted mode operation}
If the vertical clock phases are held more than about 7V below substrate  then the surface potential of the silicon underlying the phases becomes pinned at the substrate voltage (so-called Inverted-mode operation or IMO). This has important consequences for both dark current and CIC. If during the read-out of the image the clock phases never become inverted (so called Non-inverted mode operation or NIMO) then the CIC is greatly reduced (in the case of QUCAM2 it fell from 0.2e$^{-}$ pixel$^{-1}$ to a level that was unmeasurable even after 50 bias frames had been summed). At higher operational temperatures NIMO will cause an approximate 100-fold increase in dark current: something that will negate any gains from lowered CIC. The trade-off between CIC and dark current does not, however, hold at lower temperatures. For QUCAM2 at an operational temperature of 178K, the NIMO dark current was 1.5e$^-$pixel$^{-1}$hour$^{-1}$, a value that was found to be constant for exposure times of up to 1000s. Switching to IMO reduced this to $\sim$0.2e$^-$pixel$^{-1}$hour$^{-1}$: a value somewhat difficult to measure since it is $\sim$4 times below the current delivered by cosmic ray events. This demonstrates that operation at cryogenic temperatures permits NIMO  without a loss of performance from high dark current. It should be noted that at higher temperatures, such as those experienced by Peltier-cooled CCDs, the situation becomes more complex since dark current is no longer constant with exposure time. This effect was seen during the optimisation of QUCAM2 when operated experimentally at 193K. Using NIMO, the dark current for 60s exposures was measured at 12e$^-$pixel$^{-1}$hour$^{-1}$ whereas for 600s exposures it was 40e$^-$pixel$^{-1}$hour$^{-1}$.
%%%%%%%%%%%%%%%%%%%%%%%%%%%%%%%%%%%%%%%%%%%%%%
\section{Modes of EMCCD operation}
\label{sec:modes}
EMCCDs can be utilised in three separate modes, each offering optimum SNR in certain observational regimes. In this section the equations describing the SNR in these three modes are shown.
\subsection{Conventional mode}
The SNR obtained through the conventional low-noise amplifier is given by:
\begin{equation}
\text{SNR}_{C}=\frac{M}{\sqrt{M+\nu_C+D+K+\sigma_N^2}}{  },
\label{eq:snrconventional}
\end{equation}
where $M$ is the mean signal per pixel from the source, $\sigma_N$ is the read noise from the conventional amplifier,  $\nu_C$ the mean CICIR per pixel, $D$ the dark charge and $K$ the charge received from sky photons.
\subsection{Linear mode}
In linear mode, the digitised signal from the EM output is interpreted as having a linear relationship with the photo-electrons, as is usual for a CCD. The SNR obtained through the EM output is then given by:
\begin{equation}
\text{SNR}_{lin}=\frac{M}{\sqrt{2.(M+\nu_C+D+K)+(\sigma_{EM}/g_A)^2}}{  }.
\label{eq:finalLIN_SNR}
\end{equation}
The factor of 2 in the denominator accounts for the multiplication noise (see section 2.2). The derivation of this factor can be found in
\cite{MarshHTRA} and \cite{TubbsPhD}. The read noise in the EM amplifier $\sigma_{EM}$ will typically be tens of electrons due to its higher bandwidth but its contribution to the denominator is rendered negligible by the use of high EM gain, g$_A$.
High speed means high read noise so high frame rate cameras will need higher gains than the more leisurely QUCAM2 (1.6s read-out time in EM mode).

\subsection{Photon counting mode}
\label{sec:pcsnr}
\begin{figure}
\begin{center}
\includegraphics[width=0.5\textwidth]{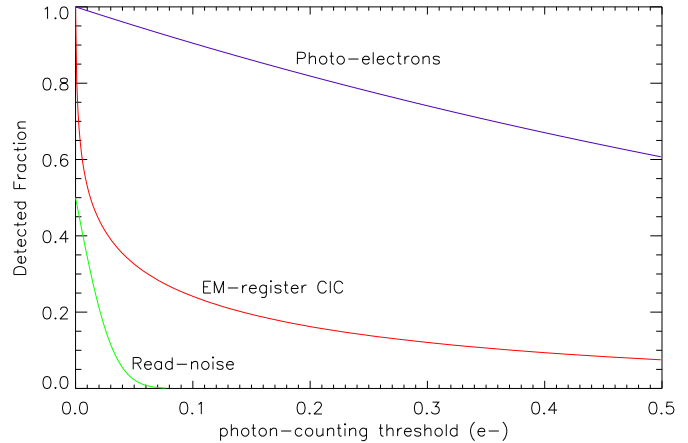}
%generated by "CalculateSNRPC_ALL"
\caption[Threshold]{The effect of the PC threshold value on the detected
  fraction of the signal, the CIC and the read noise in an EMCCD image. In
  this case the read noise $\sigma_{EM}=0.025e_{pe}^-$ and $g_A=2000$.}
\label{fig:PCthreshplot1}
\end{center}
\end{figure}

In PC mode we apply a threshold to the image and interpret any pixels above it as containing a single photo-electron. This leads to coincidence losses at higher signal levels where there is a significant probability of a pixel receiving two or more photo-electrons, but at weak signal-levels where this probability is low, PC operation offers a means of eliminating multiplication noise (\citealt{Plakhotnik}) and obtaining an SNR very close to that of an ideal detector. When photon counting we must aim to maximise the fraction of genuine photo-electrons that are detected whilst at the same time minimising the number of detected CICIR and read noise pixels. Figure \ref{fig:PCthreshplot1} shows how this can be done.
The graph shows us that if we set a photon-counting threshold of, say, 0.1e$_{pe}^-$ (a value that was later found to be optimum, see Section~\ref{sec:threshchoice}) we will detect 90\% of photo-electrons but only 23\% of the CICIR. False counts from the read noise will be negligible.

The SNR of an ideal photon-counting detector, including the effects of coincidence losses, is given by:
\begin{equation}
\text{SNR}_{pc}=\frac{M}{\sqrt{e^M-1}}{  }.
\label{eq:SNRPC_IDEAL}
\end{equation}
This equation (derived in Appendix~\ref{sec:PCnoiseDerivation}) needs to be modified to accurately describe a photon-counting EMCCD since it makes no allowance for the complex effects of CICIR and choice of threshold level. Since the distribution of CICIR and photo-electron events are different, the SNR can vary greatly depending on the precise choice of threshold. These complexities have been explored  in detail by modelling (see Section~\ref{sec:modelsection}) but, in short, the following SNR relation (derived in Appendix \ref{sec:SNRderivationPC}) is found to hold, assuming a PC threshold of 0.1e$^-_{pe}$ (close to optimum, see Section 4.5):
\begin{equation}
\frac{0.9M}{\sqrt{\delta}\sqrt{\exp{[(0.9(M+D+K)/\delta)+0.23\ln(g_A)\nu_C]}-1}}{   }.
\label{eq:SNRPC_EMCCDblocking}
\end{equation}
It should be noted that the maximum possible SNR in a single PC frame is $\approx 0.8$ (see Figure 11) and it is then necessary to average many frames to arrive at a usable image (i.e. one with SNR $>3$). The number of frames that would need to be \textquoteleft blocked\textquoteright{ }together in this fashion is given the symbol $\delta$ in  Equation~6.
%As $\delta$ increases we can then photon count on ever increasing signal fluxes, limited only by the %camera\textquoteright s maximum frame rate.
\subsection{Optimum choice of mode}
Considered from the point of view of maximising the SNR per pixel, the choice of readout mode is quite simple. Figure~7 shows the range of per-pixel illuminations over which each mode offers the best SNR, based on the equations presented earlier in this section. A single pixel is, however, not usually the same thing as a  single wavelength element in a  reduced spectrum. Many additional factors affect the choice of mode, such as plate scale, seeing and sky background, and this is explored in greater depth in Section~\ref{sec:receta}.
\begin{figure}
% Generated by C:\MNRAS_paper2\Final_IDL\ThreeRegimes4.pro
\begin{center}
\includegraphics[width=0.5\textwidth]{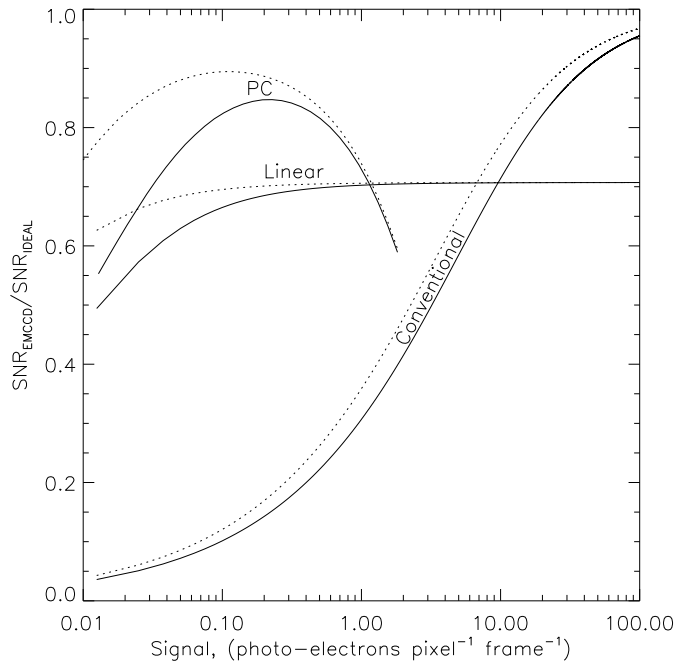}
\parbox{0.5\textwidth}{
\caption[Three Regimes]{The relative SNR (compared to an ideal noise-free detector of the same QE) of each EMCCD  mode (conventional,linear and PC) is shown over a wide range of illuminations. The solid curves show the performance of a detector identical to QUCAM2 ($\nu_C$=0.013e$^-$, $\sigma_N$=3.1e$^-$), the dotted lines shows the performance that would be expected from a more highly-optimised detector ($\nu_C$=0.003e$^-$, $\sigma_N$=2.6e$^-$).}}
\label{fig:three_regimes}
\end{center}
\end{figure}
%
%%%%%%%%%%%%%%%%%%%%%%%%%%%%%%%%%%%%%%%%%%%%%%
\section{Modelling photon-counting performance}
Modelling was required for two reasons. First there was no equation available describing the output signal distribution of pixels affected by CICIR and second to check that the assumptions underlying the derivation of Equation~\ref{eq:SNRPC_EMCCDblocking} (see Appendix~C) are valid.
\label{sec:modelsection}
\subsection{What was modelled}
Equation~\ref{eq:BasdenEquation} describes the output of an EM  register for any integer number $n$ of photo-electrons. Using the Poisson distribution it is possible to calculate the proportion of pixels that contain $n=1,2,3....$ photo-electrons as a function of the mean illumination $M$. This result can be combined with Equation~\ref{eq:BasdenEquation} to yield the output distribution of the EM register for any \textit{mean} input signal level. No equivalent relation describing the output distribution of CICIR could be found and it is here that Monte Carlo modelling is required (see Section~4.2).

The final output of the model is a pair of 3D vectors. The first of these shows the output pixel value distributions (i.e. histograms) for a wide range of signal and CICIR levels. The second is the associated cumulative distribution function (CDF) derived from the histograms in the first 3D vector. The CDF is extremely useful since it indicates the mean photon counts per-pixel that we can expect for any combination of signal and CICIR for any given PC threshold.  These 3D vectors can be visualised as two cubes of normalised histogram values and CDF values. The $x$-axis of the cubes are labelled with a logarithmic-spaced range of CIC values and the $y$-axes with a logarithmic-spaced range of signal values (extending up to a maximum of 2.5e$^-$). The $z$-axes are labelled with  pixel values extending up to a maximum of 10e$_{pe}^-$. In the case of the CDF cube, the $z$-axis units can also be interpreted as the PC threshold setting and the data values as being the mean per pixel-PC signal at that given threshold.
To illustrate what we mean, we show some sample data taken from the first of these 3D vectors in Figure~E1.

\subsection{Modelling of CICIR}
Since no analytical formula describing the distribution of CICIR events could be found it was necessary to do a Monte Carlo model of the EM register. Binomial statistics describe processes whose final outcome depends on a series of decisions each of which has two possible outcomes. It therefore applies to the creation of CIC as a pixel is clocked along the EM register. Synthetic images were generated containing between 1 and 6 CIC events per pixel. It was not really necessary to go any higher than 6 CIC events since the binomial distribution shows that there is an insignificant probability of more than 6 events being generated per pixel for mean CIC event levels of up to 0.3 per pixel, well beyond the useful operational range of an EMCCD (indeed QUCAM2 gave $\sim0.08$ CIC events per pixel). Histograms of these 6 images were then calculated to yield a set of output CIC distributions for integer input, analogous to the distributions for photo-electron events given by Equation~\ref{eq:BasdenEquation}. It was then necessary to combine these histograms using the binomial distribution formula to yield the output distribution of the EM register for any \textit{mean} CIC event level.

The model was implemented by simulating the transfer of charge through an EM register with 604 elements, one pixel at a time. At each pixel transfer a dice was thrown for each electron in the pixel to decide if a multiplication event occurred. An overall EM gain of $g_A$=2000 was used. Six thousand lines were read out in this way to get a good statistical sample of pixel values.  At the start of the readout of each simulated image row, the EM register was charged with a single electron per element. This simply amounted to initialising the array representing the EM register with each element equal to 1.  This was then read out, simulating the effect of  charge amplification, to yield an image with width equal to the length of the register. The resulting image was then scrambled (i.e. the pixels were reordered in a random fashion) and added to its original self to yield an image containing an average of 2 CIC events per pixel. This scrambling was necessary since the raw images contained pixels with values that were approximately proportional to their column coordinate, with the pixels in the higher column numbers having higher values. Further scramble-plus-addition operations were  performed to yield  images  with  3, 4, 5 and 6 events per pixel.

\subsection{Modelling of realistic EMCCD images}
The distributions of the CICIR (calculated in Section 4.2) and the distribution of the photo-electron events  were then combined, together with read noise, through the use of intermediate model images. Photo-electron events were first generated in the image using a random number generator that was weighted by the  distribution of the EM register output. CICIR events were then added to the image in the same manner as for the photoelectrons. Finally, read noise of $\sigma_{EM}=0.025$e$_{pe}^-$ was added to every pixel in the image. These model images contained a bias region from which photo-electrons were excluded. Histograms of the image and bias regions were then calculated to yield the distributions and their CDFs. These CDFs effectively gave the mean photon-counting signal from the image and bias areas as a function of $t$ the PC threshold. The SNR could then be determined using the following equation (derived in Appendix~\ref{sec:PCnoiseDerivation}):
\begin{equation}
\text{SNR}_{pc}(t)=\frac{-\ln[1-CDF_I(t)]+\ln[1-CDF_B(t)]}{\sqrt{[1-CDF_I(t)]^{-1}-1}}{ },
\label{eq:neweq35}
\end{equation}
where $CDF_I$ and $CDF_B$ are the image and bias area CDFs, respectively. Since the CDFs were calculated over a wide range of threshold values it was possible to find the optimum threshold value or alternatively just calculate the SNR for any given  threshold.
\subsection{Testing of the model}
The model output was tested against a stack of 45 QUCAM2 bias frames of known CIC level, EM gain  and system gain. The comparison between model and data is shown in Figure~8. The agreement is good, although QUCAM2 shows a slight excess of low value events compared to that predicted. This can be explained by the imperfect charge transfer in the EM register which boosts the relative number of low-value events, an effect that was not included in the model.
\begin{figure}
% Generated by C:\MNRAS_paper2\Final_IDL\testofModelQUCAM2.pro
\begin{center}
\includegraphics[width=0.45\textwidth]{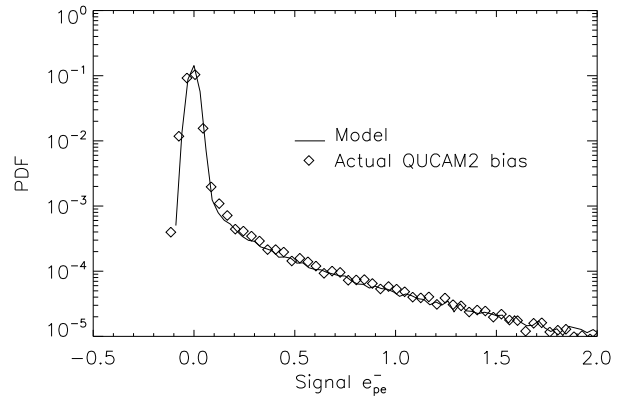}
\parbox{0.45\textwidth}{
\caption[QUCAM2 thresh]{The plot compares the histogram of pixels in a genuine QUCAM2 EMCCD image (diamonds) with that of a model image (solid line). }}
\label{fig:proofOfModel}
\end{center}
\end{figure}
\subsection{Optimum PC threshold}
\label{sec:threshchoice}
It is the  read noise that sets the lower limit on the photon-counting threshold. Gaussian statistics predict that a threshold set $\sim3\sigma$ above this noise gives 1 false count per 1000 pixels, falling to 1 pixel in 32000 if we choose a threshold of $4\sigma$. Other groups (\citealt{2006SPIE.6276E..42D}, \citealt{IvesUSPEC}) have chosen quite high thresholds (5-5.5$\sigma$). This is a good choice as long as it is combined with a high EM gain, high enough to ensure that the mean level of a photo-electron is at least 10 times the threshold value. This high ratio between mean photo-electron level and read noise permits a threshold to be set low enough to include a majority of photo-electrons. There is a limit to how  high the EM gain should be pushed, though,  since it can risk damage to the chip if gain is applied during overexposure for long periods.

Figure 9 shows  that the PC threshold needs to be tuned depending on the signal level, so that the maximum number of genuine photo-electrons are counted and the maximum number of CICIR are rejected. Figure 9 also shows that in the case of a detector with read noise $\sigma_{EM}=0.025$e$_{pe}^-$, the optimum threshold falls as low as $3.2\sigma_{EM}$. From a data-reduction point of view, using a variable threshold adds complexity but may be necessary to extract maximum SNR, particularly at low signal levels. One example of this would be the measurement of a faint emission line, the peak of which would be placed at an optimum signal level through a suitable choice of frame rate. Here, the threshold would be set low, in the region of $0.1$e$_{pe}^-$ according to Figure~9. The wings of this same line, which may be an order of magnitude fainter would then benefit from an increased threshold, say $\sim0.25$e$_{pe}^-$. The use of an adaptive threshold would create many side effects, such as noise artifacts (the amount of background signal from CIC, sky and dark current would be modified depending on threshold setting), so would require extra data reduction effort.
 Note also that non-Gaussian pattern noise is a particular problem in high-speed detectors in an observatory environment which may require the threshold to be pushed higher than would otherwise be optimum.
\begin{figure}
% Generated by C:\MNRAS_paper2\Final_IDL\CalculateSNRPC_ALL_ANALOGUE_MIX.pro
\begin{center}
\includegraphics[width=0.45\textwidth]{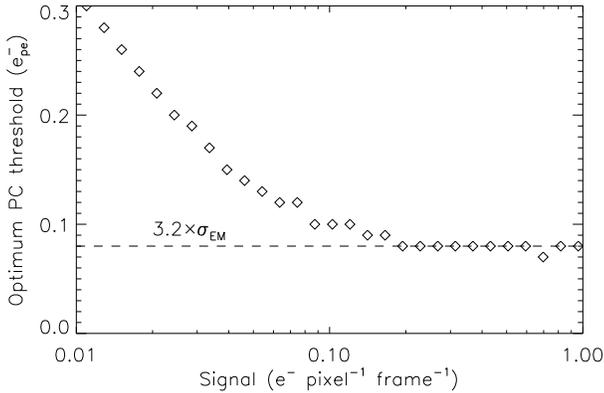}
\parbox{0.45\textwidth}{
\caption[QUCAM2 thresh]{The PC threshold that gives maximum SNR is plotted as a function of signal level. Model images were used with parameters close to those of QUCAM2. The read noise $\sigma_{EM}$ was 0.025e$_{pe}^-$ and the multiplication gain $g_A$ was 2000.}}
\label{fig:qucam2thresh}
\end{center}
\end{figure}
The models described so far have assumed a read noise of 0.025e$_{pe}^-$ (equal to that of QUCAM2). A simulation was performed of the effect of higher (0.05e$_{pe}^-$) and lower (0.012e$_{pe}^-$) read noise on the SNR performance of an EMCCD. Whilst the higher noise definitely impinges on the photon-counting performance by forcing the threshold higher and giving a lower detected fraction of photo-electrons, the lower noise gives very little additional benefits.  This can  be explained by the fact that the bulk of the CICIR has a distribution that falls between 0-0.05e$_{pe}^-$ and it will dominate any read noise lying within the same range.

In the specific case of QUCAM2 there is an additional reason (apart from staying above the read noise) why the threshold must be kept slightly elevated. This is the effect of poor charge transfer efficiency (CTE) in the EM register. Figure~10 shows the autocorrelation of a low-level flat-field  which demonstrates the problem. The autocorrelation was performed along an axis parallel to the serial register. The slight broadening of the autocorrelation peak is indicative of less-than-perfect CTE. Using a threshold much below 0.1e$_{pe}^-$ would cause complex effects from multiple counting of single electron events due to the slight tail on each event being above the threshold.
\begin{figure}
% Generated by C:\MNRAS_paper2\Final_IDL\GenerateAutoCorrelateGraphs.pro
\begin{center}
\includegraphics[width=0.35\textwidth]{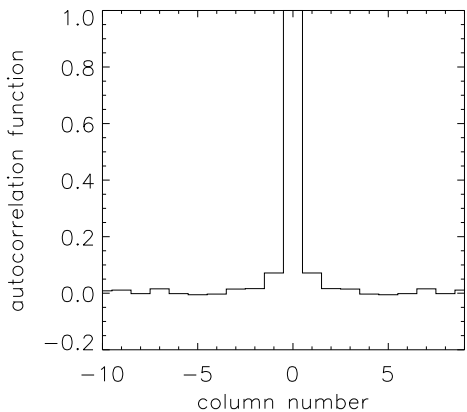}
\parbox{0.4\textwidth}{
\caption[CTE problems]{The autocorrelation of a weakly illuminated QUCAM2 image showing the elongation of the single electron events through CTE degradation in the EM register. }}
\label{fig:poorCTE}
\end{center}
\end{figure}

Note that \cite{PCStrategies} have modelled a multiple threshold technique that can extend photon-counting operation well into the coincidence-loss dominated signal regime (see Section~3.3) whilst maintaining high performance.
\subsection{Simplification of SNR equation in PC mode}
\begin{figure}
% Generated by C:\MNRAS_paper2\Final_IDL\TestOfApproximation.pro
\begin{center}
\includegraphics[width=0.45\textwidth]{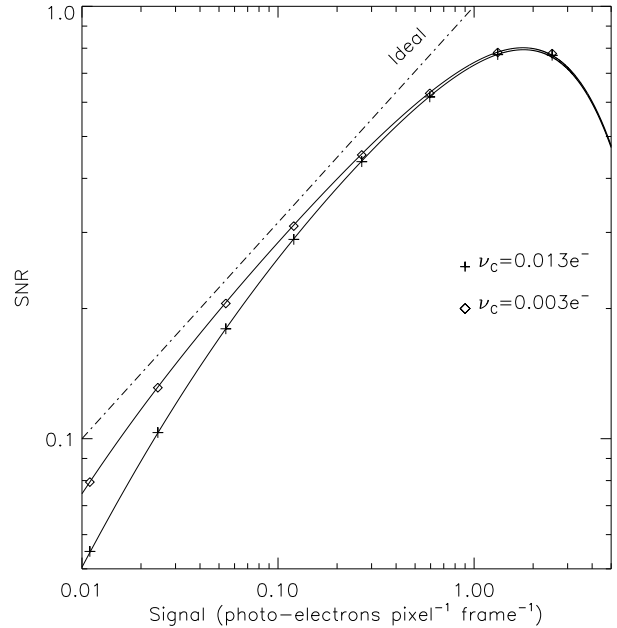}
\parbox{0.45\textwidth}{
\caption[Model Versus Equation]{The SNR of two hypothetical EMCCDs in photon-counting mode, one with CICIR $\nu_C$ equal to that of QUCAM2, another with $\nu_C$ equal to that which might be obtained in a more optimised detector. $g_A$ in both cases is 2000 and the read noise is 0.025e$_{pe}^-$. The crosses and stars show the predictions of the Monte Carlo model (Equation~7), the solid lines show the approximation (Equation~\ref{eq:SNRPC_EMCCDblocking}). The SNR of an ideal detector (one where the SNR$=\sqrt{\text{Signal}}$) is plotted as a dot-dash line for comparison.}}
\label{fig:modelVequation}
\end{center}
\end{figure}
The method used to calculate SNR in PC mode (Equation~\ref{eq:neweq35}) is rather complex since it requires the analysis of both image and bias areas in a large model image. A simpler SNR formula that can be applied more generally was therefore sought. One simplification would be to consider CIC, read noise and photo-electrons separately, i.e. assume that they only interact in the digital domain after thresholding. This is, of course, an approximation and in reality there is a complex interplay between CIC events and photo-electron events in the analogue domain. For example, a photo-electron event could be \textquoteleft helped over\textquoteright{ }the PC threshold by an accompanying CIC event, or a CIC event could be lost by occurring within an illuminated pixel. A second simplification would be to assume that, as the signal or CIC  level increases, the detected fraction of these events does not change. It is thought reasonable to make these approximations since it will only seriously fail in the high signal regime where there is a  high probability of coincidence losses and consequently low SNR compared to an ideal detector. The resulting simplified SNR equation (Equation~\ref{eq:SNRPC_EMCCDblocking}) is derived in Appendix~\ref{sec:SNRderivationPC}.
This approximation was tested against the earlier more comprehensive model (i.e. that which used Equation \ref{eq:neweq35}) for a whole range of signals and at two CIC levels. The comparison is shown in Figure~11. As can be seen, there is excellent agreement, justifying our simplifications.

\subsection{SNR predictions from model}
\label{sec:snrgraphs}
The model was used to evaluate the photon-counting SNR (SNR$_{pc}$) over a range of CIC and signal levels. The results are  shown in Figure~12, expressed both as a fraction of the SNR of an ideal detector SNR$_{ideal}$ and the SNR of an EMCCD operated in linear mode SNR$_{lin}$  (as described in Equation~\ref{eq:finalLIN_SNR}). The ideal detector is assumed to have the same QE as the EMCCD but does not suffer from read noise, coincidence or threshold losses. The figures reveal the presence of a photon counting \textquoteleft sweet-spot\textquoteright, where an SNR in excess of 90\% of ideal is possible if the CIC can be sufficiently reduced (to around $\nu_C=0.002\,$e$^-$ pix$^{-1}$). The sweet-spot is quite narrow and extends  between signals of $\approx0.07$ and 0.2 e$^-$ pix$^{-1}$. The plots also show that for signals of less than 1.2e$^-$ pix$^{-1}$, photon counting is superior to linear-mode operation, and this is fairly independent of CIC level.
\begin{figure*}
% Generated by C:\MNRAS_paper2\Final_IDL\CalculateSNRPC_ALL_APPROXIMATION.pro
\begin{center}
\includegraphics[width=0.95\textwidth]{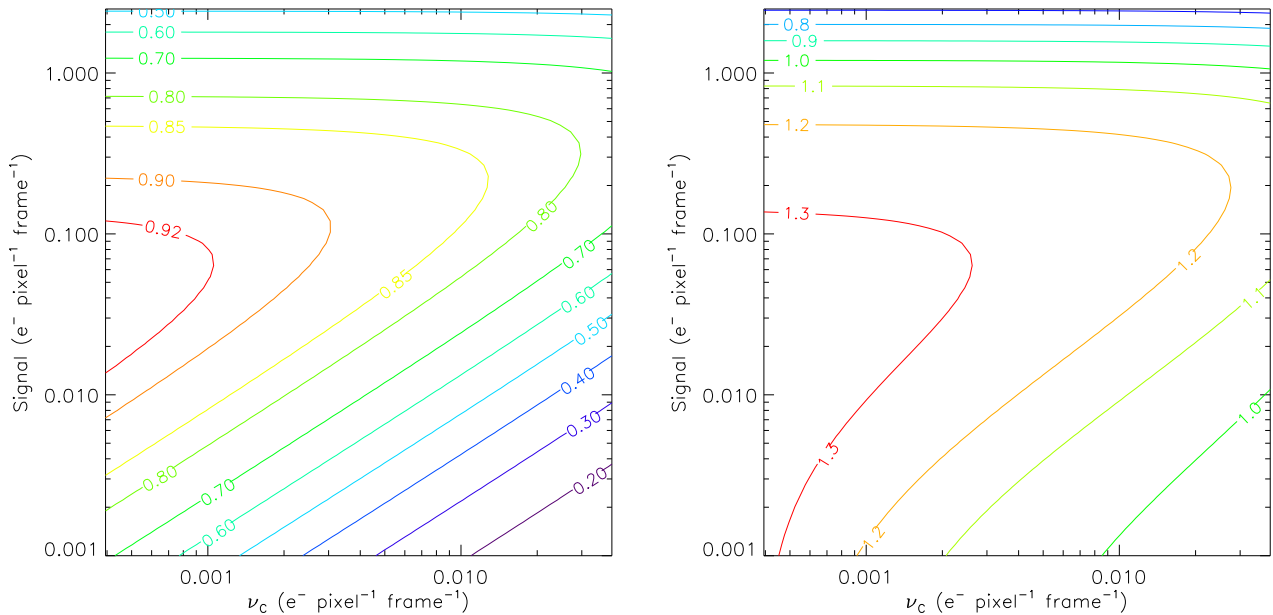}
\parbox{0.95\textwidth}{
\caption[Photon-counting SNR]{Relative SNR contours of a photon-counting EMCCD compared in the left panel to an ideal detector and in the right panel to a linear-mode EMCCD.   It has been assumed that all the CIC is generated in the EM register, the read noise $\sigma_{EM}$=0.025e$_{pe}^-$, the threshold is fixed at 0.1e$_{pe}^-$ and  the EM gain, $g_A$=2000.}}
\label{fig:PCvsBOTHplot1}
\end{center}
\end{figure*}

Figure~12 shows that the sweet-spot of an EMCCD actually covers a very small range of
exposure levels, however, we can effectively slide the sweet-spot along the
per-temporal-bin exposure scale to quite high signal levels through the use
of blocking, i.e. summing together a number of frames whose total exposure time equals our required temporal resolution.
%This is demonstrated in Figure~15.
For QUCAM2, if we use
the fairly generous definition of the sweet-spot as occupying the
exposure range over which SNR$_{pc}>\,$SNR$_{lin}$ then this dynamic
range is about 30:1. This would be like using a normal science CCD
camera with a 7-bit (16-bit being more usual) analogue to digital converter (ADC) and could cause problems if the spectrum we
wish to observe has a set of line intensities that exceeds this
range.
%%%%%%%%%%%%%%%%%%%%%%%%%%%%%%%%%%%%%%%%%%%%%%
\section{Recipe for using an EMCCD  for astronomical spectroscopy}
\label{sec:receta}
The QUCAM2 camera at the WHT is used here as the basis of an example of how to correctly set up an EMCCD to maximise the SNR of a spectroscopic observation. The general principles outlined here, however, apply to any EMCCD camera.

QUCAM2 is a relatively slow camera giving a minimum frame time in EM mode of 1.6s. This then dictates the highest temporal resolution that is available.  The small size of the CCD means that when used on the ISIS spectrograph it measures only 3.3 arc-minutes in the spatial direction. Full frame readout is then generally needed in order to locate a suitable comparison star for slit-loss correction and this frame rate will therefore be hard to improve upon through the use of windowing. The linear EM mode should be considered as the default mode about which the observations are planned. The reason for this is that it gives an SNR that is a constant fraction of an ideal detector for almost all signal levels (see Figure~7). The observer will not go far wrong by selecting this mode. It may be possible  to coax extra SNR (as much as 40\%) from the observations by switching to one of the other modes but if the observation is not prepared carefully the data could prove useless. With linear mode the observer is guaranteed a practically noise-free detector without the dangers of potential coincidence losses and worse SNR than with a conventional CCD.

The first stage in planning the EMCCD observation is to refer to Figure~13. This allows us to calculate $m$, the flux per pixel step in wavelength that we can expect from the object at our chosen spectral resolution.
\begin{figure*}
% Generated by C:\MNRAS_paper2\Final_IDL\Gratings.pro
\begin{center}
\includegraphics[width=0.75\textwidth]{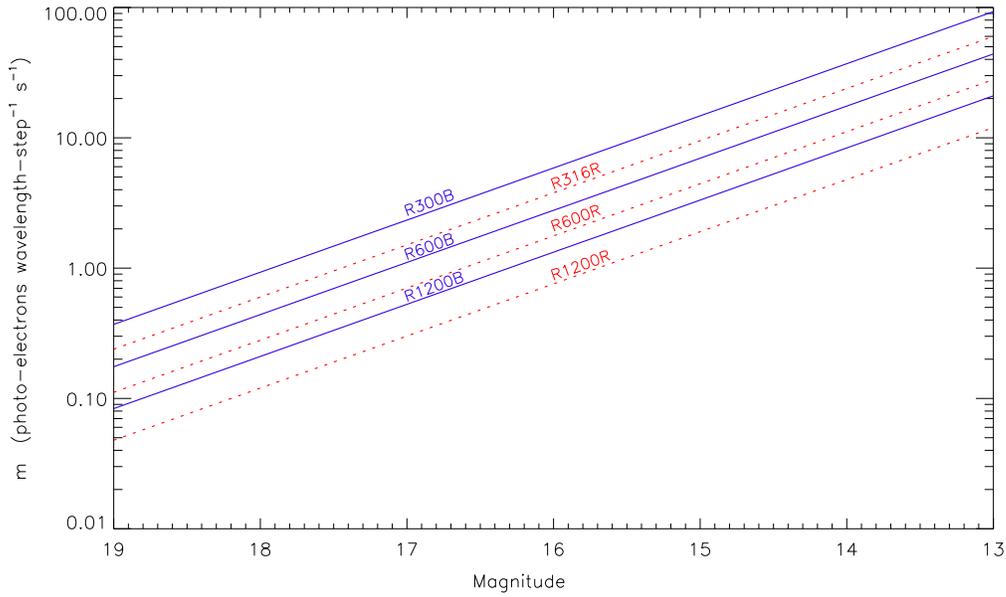}
\parbox{0.75\textwidth}{
\caption[WHT gratings]{The flux $m$, that can be expected from each grating on ISIS when used with QUCAM2 as a function of source magnitude. $R$-band magnitudes as shown for the $R$-gratings and $B$-band magnitudes for the $B$-gratings. Calculated for airmass=1.}}
\label{fig:gratings}
\end{center}
\end{figure*}
Using this datum we then need to refer to Figure 14 to see how many seconds of observation ($T_{\text{SNR1}}$) would be required on our object, using an EM detector in linear mode, to reach an SNR=1 in the final extracted spectrum.  Figure 14 shows the calculated times for both dark and bright-sky conditions (new and full moon) for both arms of ISIS. Seeing of 0.7" and a slit width of 1" has been assumed. Given that the spatial plate-scale of QUCAM2 on ISIS is 0.2" pix$^{-1}$, this implies that each wavelength element in the final extracted spectrum contains $\sim5$ pixels-worth of sky (assuming the spectrum is extracted across $\sim1.5\times$FWHM pixels in the spatial direction). Once we know the time it takes to reach an SNR=1, it is then straightforward to calculate the time needed to reach any arbitrary SNR, since with linear mode, the SNR is proportional to the square-root of the observation time.
\begin{figure*}
% Generated by C:\MNRAS_paper2\Final_IDL\Sigmas1.pro
\begin{center}
\includegraphics[width=0.95\textwidth]{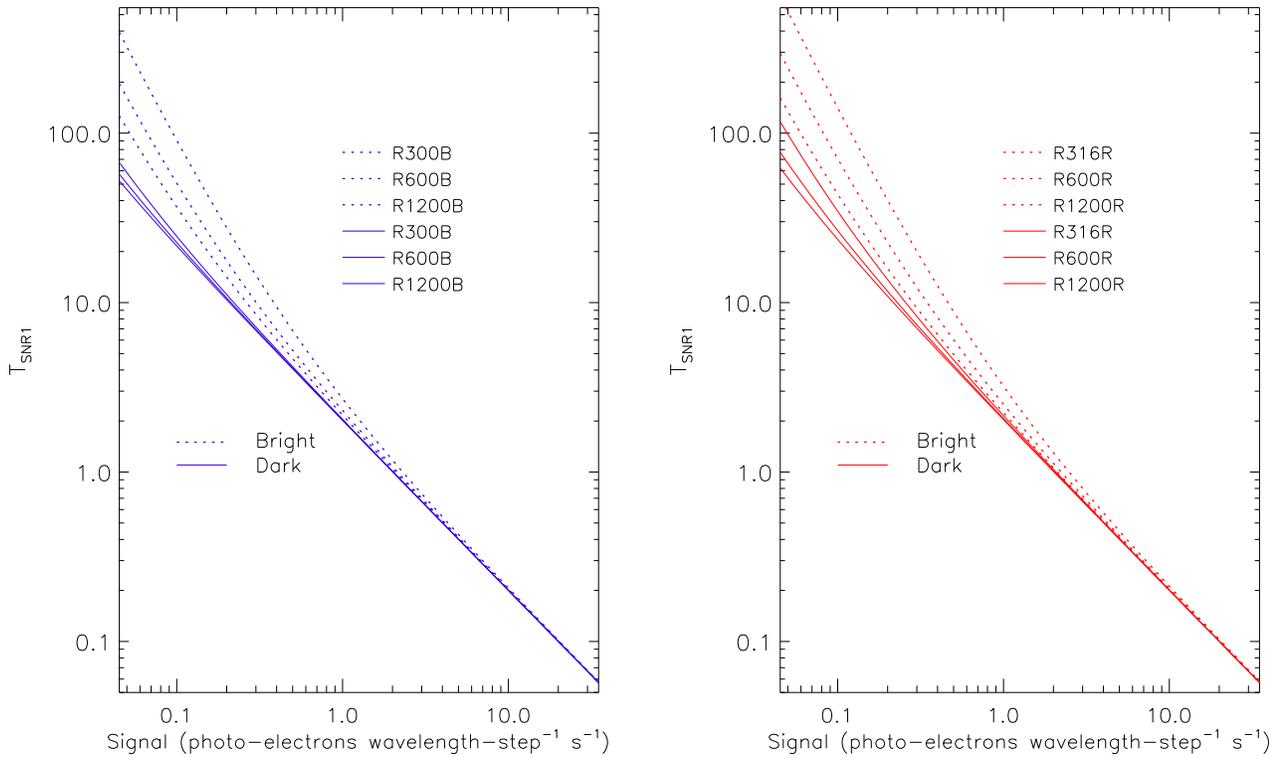}
\parbox{0.95\textwidth}{
\caption[WHT gratings]{$T_{\text{SNR1}}$, the observation time on the WHT required to reach an SNR of 1 in the final extracted spectrum with the ISIS blue (left panel) and red (right panel) gratings as a function of source brightness. The detector is QUCAM2 operated in linear mode.  A slit-width of 1",  seeing of 0.7" and a spectral extraction over  5 pixels in the spatial direction are assumed. Observations are in the B-band (left) and $R$-band (right).}}
\label{fig:sigmasBoth}
\end{center}
\end{figure*}

The observer can now either play safe and use linear mode or explore the possibility of up to 40\% higher performance from either PC or conventional modes. This will depend on the mean per-pixel signal (from all sources including the sky) that we can expect during an exposure of duration equal to our required temporal resolution, $\tau$. This is calculated as follows:
\begin{equation}
\text{signal pix}^{-1}=\tau .\left[\text{sky}+\frac{m}{d}\right],
\label{eq:findPixSig}
\end{equation}
where $d$ is the seeing-induced FWHM of the spectrum along the spatial axis of the CCD frame, measured in pixels. This dictates what spatial-binning factor we later need to use when extracting the spectrum.
Note that in conventional and linear mode, $\tau$ is defined by the individual frame time, whereas in PC mode we divide $\tau$ into $\delta$ separate frames that are later photon counted and summed (in order to observe brighter objects without incurring coincidence losses). Now that we have an estimation of the per-pixel signal we can  use Figure 15 to find which mode will give the optimum SNR on a per-pixel basis.
\begin{figure*}
% Generated by C:\MNRAS_paper2\Final_IDL\SelectMode.pro
\begin{center}
\includegraphics[width=0.75\textwidth]{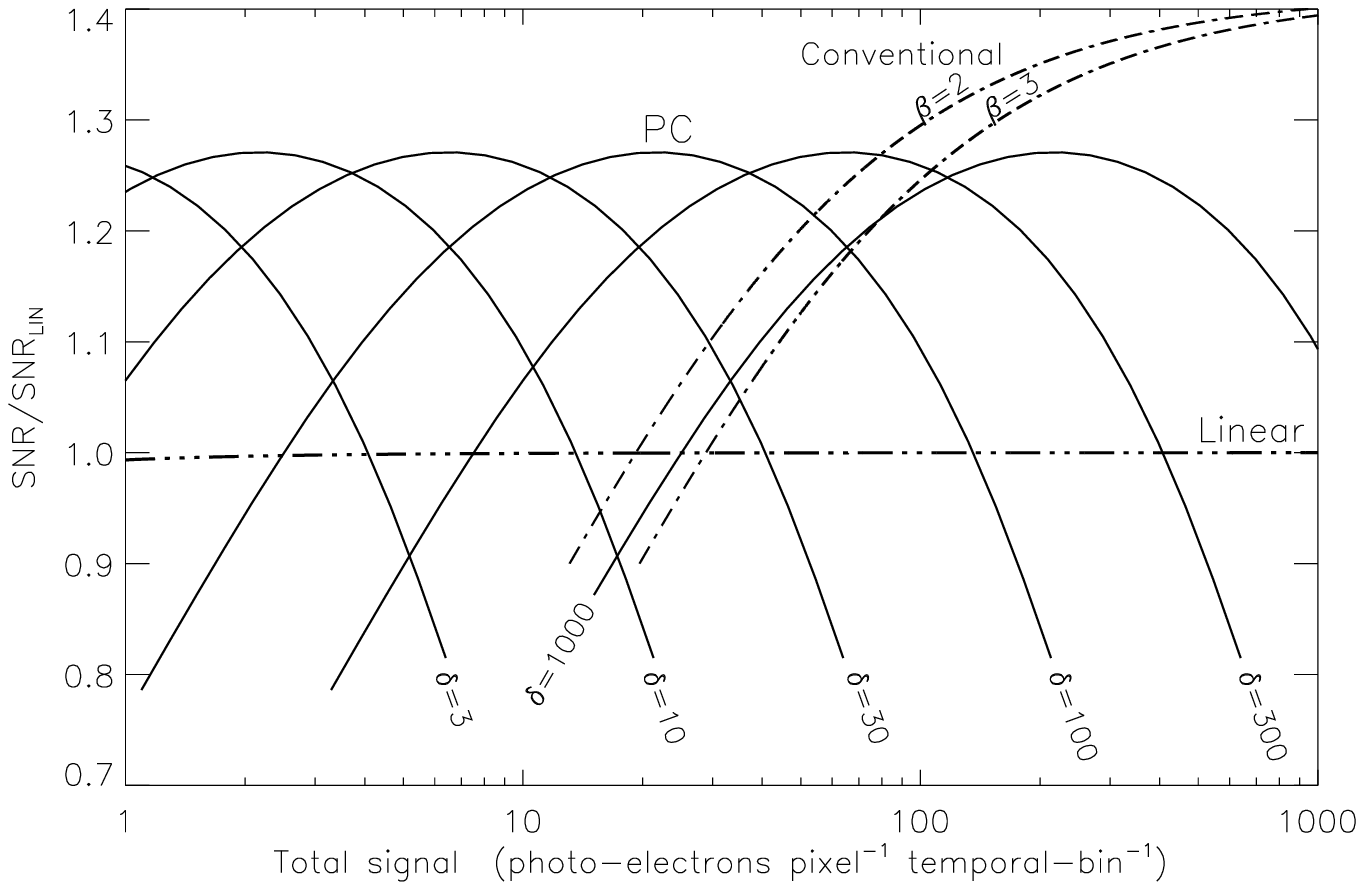}
\parbox{0.95\textwidth}{
\caption[1 temp bin]{The relative SNR achievable using each mode as a function of the total (i.e. source+sky) per-pixel signal. $\beta$ is the off-chip binning factor. $\delta$ is the photon counting blocking factor (the number of thresholded PC frames that are summed within each temporal bin). The read noise and CIC levels experienced with QUCAM2 are assumed.}}
\label{fig:onetempbin}
\end{center}
\end{figure*}
If we find we can use photon counting then, according to Figure 15, the SNR gain will be $\sim25$\%, which we can translate into less telescope time. The SNR increases as the square root of the observation time in PC and linear mode, since the detector is dominated by noise sources with variances that increase linearly with signal. If instead we find that we have enough photo-electrons to permit conventional mode, the reduction in the observation time is harder to estimate since the relatively high read noise gives a non-linear relation between the square root of the exposure time and the SNR. The saving in telescope time from use of conventional mode could, however, be as much as 50\% (relative to linear mode) as the per-pixel signal tends to higher and higher values and the  read noise becomes insignificant relative to the Poissonian noise in the sky and target.

\subsection{Worked example of the recipe}
We wish to observe an R=18.5 eclipsing binary star with the red arm of ISIS at the highest possible spectral resolution under dark-sky conditions. The emission line is $4\times$ brighter than the underlying continuum. The star undergoes an eclipse that lasts 7 minutes that we wish to resolve spectroscopically. What SNR can we achieve? Which mode should we use?

To begin this problem, we use Figure~13 to establish  the signal that we can expect from the emission line. If we choose the R1200R grating (resolution $\sim7000$) we will receive $0.07\times4=0.28$ photo-electrons per wavelength step per second. If we observe the object with a temporal resolution of 30s we will be able to easily resolve the eclipse. Referring to Figure~14 we can then see that this combination of signal and spectral resolution will require $\sim7$s of observation to give an SNR=1 in the final extracted spectrum if we use linear mode. Since we can actually observe for 30s, the SNR will be equal to $\sqrt{30/7}=2.1$.

We now turn to the choice of observing mode. Assuming 0.7" seeing, a slit-width of 1" and that the spectrum will be extracted over 5 pixels in the spatial direction, we can calculate that the peak signal per temporal bin per pixel will be $0.28\times30/5=1.7$ e$^-$. To this we must add the sky signal tabulated in Table~\ref{tab:dark_tableWHT}. Since we are observing in dark time and at high resolution this will be a negligible 0.003e$^-$s$^{-1}$. Referring to Figure~15 we can then immediately rule out conventional mode as the SNR would collapse at such a low signal level. The best mode would then be PC with $\delta$ somewhere between 3 and 10. Interpolating the figure we can estimate that $\delta \approx 7$ would be optimum. This is feasible since the minimum read-time for 7 PC frames is 11.2s with QUCAM2, well below the required temporal resolution of 30s. In conclusion, we would obtain the best SNR by observing at a frame time of 4.3s, photon counting the raw images and then averaging them into groups of 7 to obtain our required time resolution. Tuning the exposure time in this way to keep the spectral line on  the PC sweet-spot affords us a $\sim25\%$ SNR improvement over linear-mode operation (see Figure 15).

Note that there is an upper-limit to the useful PC blocking factor $\delta$,
set by the read noise of the conventional amplifier. Large blocking
factors imply low-temporal resolution and there comes a point where it
becomes favorable to use the conventional mode (see Figure~15). The
degree of off-chip binning $\beta$ used is critical since this adds
noise to conventional mode but not to photon counting mode.

In the case of QUCAM2 it can be demonstrated (using the SNR equations in
Section \ref{sec:modes}) that the maximum useful blocking factor is
$\sim20\sigma_N^2\beta$.  For larger values of $\delta$ the SNR that can be obtained with conventional mode then exceeds the maximum obtainable with PC operation.
\subsection{Binning}
CCDs can be binned on-chip in a noiseless fashion in both axes. Since a spectrum will always be spread by seeing in the spatial direction, some degree of binning is often required. Some of this can be done on-chip but it is usual to do some of it off-chip (i.e. post-readout)  during extraction of the spectrum.
An EMCCD will suffer much less from off-chip binning than a conventional detector, which has an effective read noise  multiplied by the square-root of the binning factor. As the off-chip binning factor $\beta$ increases, the balance tips ever more in favour of the EMCCD.
\subsection{Phase folding}
For observations of objects that vary on a regular period, such as short-period binary stars, we can also consider extending our observations over many orbits and then \textquoteleft phase folding\textquoteright{ }the data. This is an equivalent form of off-chip binning. It increases our signal by the folding factor $\digamma$ and permits us to observe fainter sources or, alternatively, to observe the same source at higher time resolution.
If the  SNR we require per wavelength element in our final extracted spectrum is given by  SNR$_{req}$ then
the number of phase folds $\digamma$ we need to use (i.e. the number of orbits we must observe) in linear mode is given by:
\begin{equation}
\digamma\approx\frac{T_{\text{SNR1}}\times\text{SNR}^2_{req}}{\tau}{  }.
\label{eq:calcFaprox}
\end{equation}
%%%%%%%%%%%%%%%%%%%%%%%%%%%%%%%%%%%%%%%%%%%%%%
\section{EMCCDs on large telescopes}
The performance of an EMCCD on an extremely  large telescope and, in particular, the quantitative advantage it might give over a conventional CCD camera is  explored in this section. Signal fluxes are calculated assuming a hypothetical  ISIS-type instrument on the E-ELT (42-m aperture) using higher efficiency volume-phase holographic (VPH) gratings. These typically offer a 30\% increase in throughput over conventional diffraction gratings. Sky backgrounds from Paranal (approximately 20 km distant from the future E-ELT site) are assumed (\citealt{Patat2003}). The QE of current EMCCDs are already very high (peaking at $>90$\%) and any future EMCCD is unlikely to be significantly different.
% A final assumption is that the  plate scale of this E-ELT spectrograph is the same as ISIS, i.e. it is a natural seeing instrument.
The temporal resolution as a function of source magnitude is calculated assuming that an SNR of 1 is required in each element of the final reduced spectrum. Equations \ref{eq:snrconventional} and \ref{eq:finalLIN_SNR} were used for these calculations. The results are shown in Figure~16, where linear and conventional modes are compared with an ideal detector.  Two seeing conditions are assumed: natural median seeing (0.6") and that which may be obtained in the $R$-band with adaptive optics (AO) correction (0.1"). With natural seeing and when observing very faint sources, long integrations are required. Under these conditions, the sky contribution becomes so high that the use of EMCCDs actually degrades the SNR through the effects of multiplication noise. If instead AO is used, the sky contribution is considerably reduced and the EMCCD operated in linear mode gives an advantage right across the range of source intensities explored in the plot.
\begin{figure*}
% Generated by C:\MNRAS_paper2\Final_IDL\E_ELT_Performance.pro
\begin{center}
\includegraphics[width=0.95\textwidth]{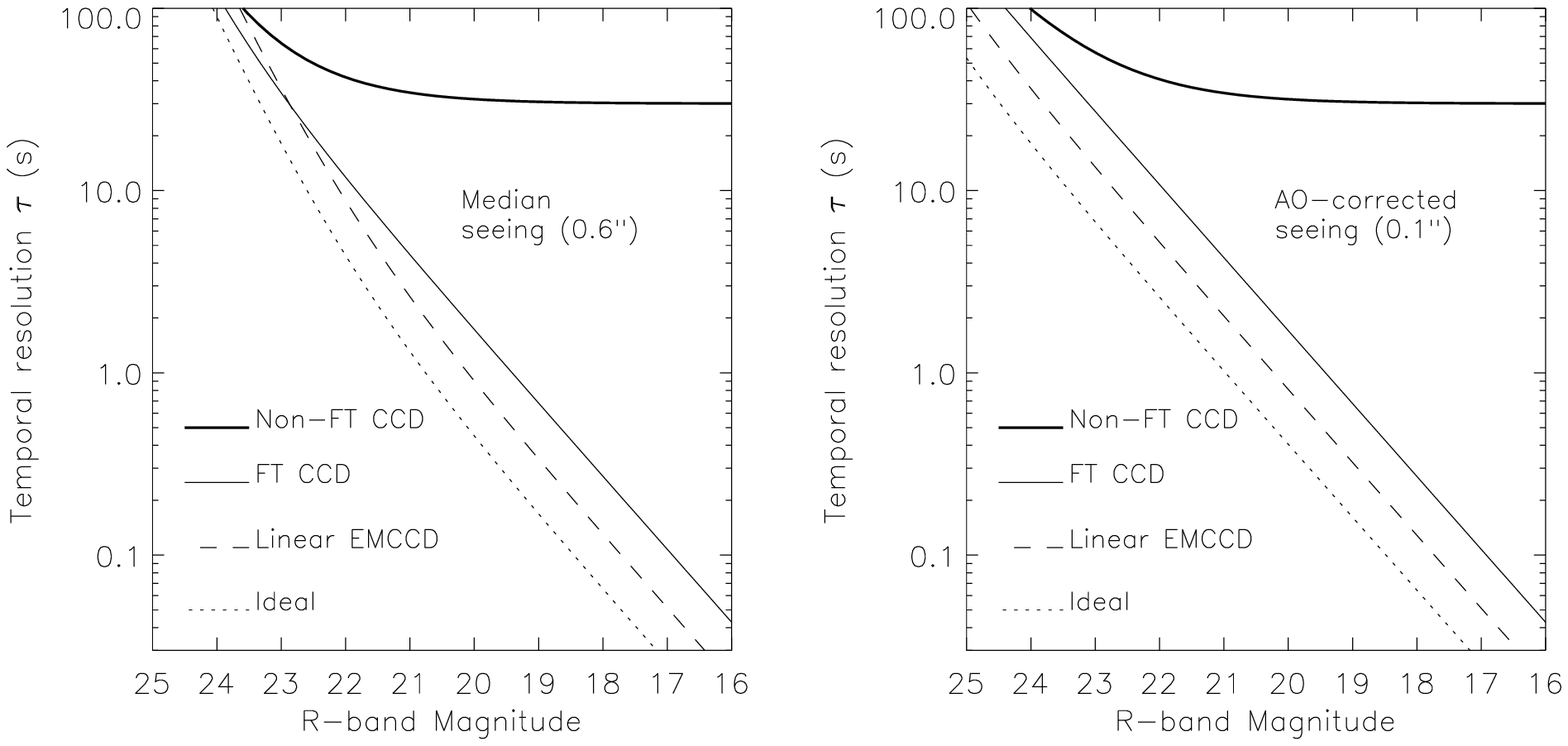}
\parbox{0.95\textwidth}{
\caption[E-ELT performance]{The temporal resolution versus source magnitude for two seeing conditions on an ISIS-type E-ELT instrument. The curves indicate the observation time required to reach an SNR of 1 in each wavelength element of the final extracted spectrum  using a
normal (non-FT) CCD, a FT CCD, an EMCCD in linear mode and an ideal
detector. Off-chip spatial binning of x2 and $\sigma_N=2.6$e$^-$ are assumed. The grating is a VPH equivalent of the ISIS R1200R grating and we observe in the $R$-band during dark time. The slit width is assumed to be 1.5x the seeing FWHM.  The curve corresponding to the non-FT CCD assumes a 2k x 4k detector (similar to that currently used on ISIS) with full-frame read out. An ideal detector is one with the same QE as a normal CCD but which has zero read noise.}}
\label{fig:ELTSNR3}
\end{center}
\end{figure*}
\subsection{Need for high frame rates}
Higher frame rates allow us to observe at higher time resolution, of course, but in the case of PC operation it also means we can avoid coincidence losses when observing brighter sources and cope with the higher sky backgrounds we can expect from the E-ELT\textquoteright s larger collecting area. We always need to ensure that the per-pixel signal remains in the sweet-spot ($<0.2$e$^-$, see Figure~7), so even in dark time and at high spectral dispersion we would need (referring to Table~\ref{tab:dark_tableELT}) to operate at $>0.5$Hz in the blue and $>$2Hz in the $R$-band if we wish to photon count on the E-ELT (natural median seeing is assumed).
\subsection{Need for frame transfer design}
Frame-transfer design reduces the dead time between exposures to a few tens of milliseconds: the time it takes to move the image into the storage area.
For a normal mechanically-shuttered non-FT camera, the dead time is equal to the read-out time of the CCD. This can have a dramatic effect on observing efficiency. For example, assuming a scientific camera is used to take blocks of 20 minute exposures for 10 hours a night, then with a read-out time of 60s this dead time could amount to more than two weeks of telescope time over the course of a year!
In high time-resolution spectroscopy with much shorter exposure times, the dead-time losses are proportionately more extreme, resulting in unfeasibly low observing efficiencies; it is here that an FT CCD can give massive gains. The performance of a conventional CCD with non-FT architecture and 30s dead time between exposures is shown plotted in Figure~16. Any future CCD used on the E-ELT, especially if destined for high time-resolution applications, be it an EMCCD or otherwise, should incorporate FT architecture.
\subsection{An EMCCD for the E-ELT}
Current EMCCDs are rather small (1k x 1k pixels). For spectroscopy this is not a good format since it limits our spectral range and also the availability of comparison stars near the target.
Any EMCCD designed for use with the E-ELT therefore needs to be physically larger. It will require multiple outputs to achieve higher frame rates, so as to permit high time-resolution and allow photon-counting mode operation. A possible geometry for such an EMCCD is shown in Figure~17. Here a monolithic 4k x 2k image area is proposed. Tapered storage areas are positioned both above and below the image area.  Their shape provides space for multiple EM registers. This is similar to the architecture of the CCD220, a smaller format, multiple output EMCCD developed for wavefront sensing (\citealt{Feautrier}).
In principle, the storage areas can be shrunk by reducing the height (in the axis perpendicular to the serial register) of their pixels. This results in more efficient use of the available silicon.  It would be at the expense of reduced full-well capacity, but this is of little concern in the low-signal regime where EMCCDs are best applied.
Each  storage area can be read out through either a conventional amplifier capable of giving 2.6e$^-$ noise at 200\,kpix s$^{-1}$ or a higher-noise (20-30e$^-$) EM amplifier capable of running at 10\,Mpix s$^{-1}$. These amplifiers could be combined so that conventional mode would be selected simply by setting the EM gain to unity, but having two separate amplifiers allows more design flexibility with one amplifier being optimised for speed and the other optimised for low noise. The target CICIR level should be around 0.003e$^{-}$. Non-inverted mode operation and liquid nitrogen cooling would ensure that dark current and CIC generated prior to the EM register are kept at negligibly low levels. Full-frame readout would be possible in 5s through the conventional amplifiers and in 0.1s through the EM amplifiers. None of these design parameters, taken individually, exceed those of current scientific cameras. We already know from discussions with E2V that such a device is feasible to manufacture, although the large pin-count may require the EM and conventional amplifiers to be combined.
\begin{figure}
% Generated by C:\MNRAS_paper2\Final_IDL\emccd-elt.odg
\begin{center}
\includegraphics[width=0.37\textwidth]{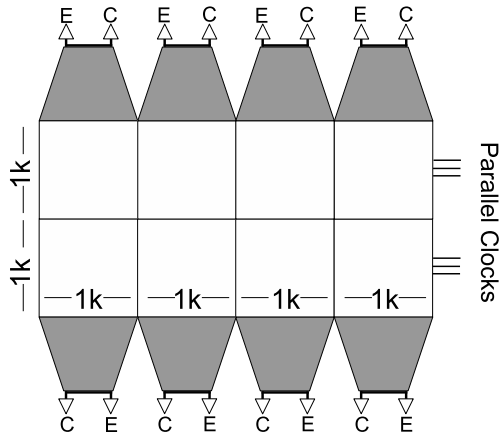}
\parbox{0.45\textwidth}{
\caption[E-ELT CCD]{A possible design for a future EMCCD for spectrographic use on the E-ELT. Conventional amplifiers are labelled C, EM amplifiers are labelled E. The shaded regions are the storage areas. Each 1k pixel x 1k pixel block has its own conventional and EM amplifiers, permitting faster parallel readout. This monolithic design is buttable along two edges allowing it to be extended horizontally for increased spectral coverage.}}
\label{fig:FutureEMCCD}
\end{center}
\end{figure}
The spectral axis would lie horizontally in Figure~17. This would maximise the spectral coverage and also allow efficient on-chip spatial binning in order to reduce the effects of CICIR and read noise, and to reduce the read time in close  proportion to the spatial binning factor. Such a CCD would have a wide application in the field of high time-resolution spectroscopy. In Figure~16 it has already been shown that simply switching to an FT design without implementing an EM register can give a huge advantage. Considering the economics of the E-ELT, the extra expense required to implement FT architecture (which approximately doubles the area of silicon required) can be easily justified. The incremental expense of then adding an EM register (estimated at 10-20\% by E2V) will give excellent value for money given the further savings in telescope time that it can provide.
\section{Conclusions}
We have shown that EMCCDs are almost perfect detectors for optical
spectroscopy. They have close to 100\% QE, virtually no read noise,
large formats, linear response, negligible dead time and are
relatively inexpensive.  Due to multiplication noise and CIC, however,
they are not as straightforward to use as conventional CCDs, and care
must be taken to ensure that they are operated in the correct mode:
conventional, linear or photon counting. If EMCCDs are used correctly,
it is possible to gain orders-of-magnitude improvement in SNR compared
to conventional CCDs at low light levels (see Figure~7), but using the
wrong mode can result in an orders-of-magnitude reduction  in  SNR. With the
aid of Monte-Carlo modelling of an EMCCD, we provide derivations of
the SNR equations for each EMCCD mode, and present a recipe for
astronomers to assist in determining the optimum EMCCD mode for a
given observation.

The EMCCD used in QUCAM2 and ULTRASPEC is the E2V CCD201-20 detector,
which at 1k$\times$1k pixels (each of $13\mu$m$\times13\mu$m) is
the largest commercially available EMCCD. However, this is still
substantially smaller than the conventional 4k$\times$2k CCDs
found on major optical spectrographs. As a consequence, using the
CCD201 results in the loss of approximately three-quarters of the
wavelength coverage and one half of the spatial coverage provided by
typical astronomical spectrographs. This factor of eight loss of
detector area is a heavy price to pay, even for the huge SNR gains of
an EMCCD. We therefore present a concept for a large-format EMCCD
which can optimally sample the focal plane of the world's major
spectrographs.  It is important to emphasize that EMCCDs are identical
to conventional CCDs in almost every respect (architecture,
performance, read-out electronics), except for the fact that they have
effectively zero read noise. This means that astronomers who do not
have read noise limited observations lose nothing by using an EMCCD
instead of a conventional CCD; in fact, we show that there will be
significant efficiency gains (equivalent to approximately two weeks of
time per year per telescope) because the frame-transfer format of EMCCDs
results in essentially zero dead
time compared to the tens-of-seconds dead time that conventional non-FT CCDs
suffer from. The biggest gains, however, will be for astronomers doing
read noise limited spectroscopy, due to the fact that the read noise
will also be zero. It is our firm belief, therefore, that once these
large-format EMCCDs become available, they will become the detector of
choice on the world's major spectrographs including those to be built for the E-ELT.

\section*{Acknowledgments}
We would like to thank Olivier Daigle for help with  derivation of the photon counting equations.
\bibliographystyle{mn2e}
\bibliography{abbrev,SMT}{}
\appendix
%\section[]{Symbols used}
%e$^-_{pe}$,{ } input referenced photo-electrons.\\
%$\nu_{C}$,{ } mean charge in a pixel from CICIR.\\
%$B_C$,{ } mean number of CICIR events in a pixel.\\
%$g_A$,{ } avalanche, multiplication or EM gain (unitless).\\
%$g_S$,{ } gain or system gain (e$^-$/ADU).  \\
%$g_{S0}$,{ } system gain of EM amplifier with EM gain=1. \\
%$p_{C}$,{ } per transfer CIC generation probability.\\
%$S$, number of stages in the EM register.\\
%$\sigma_N$,{ } conventional output read noise.\\
%$\sigma_{EM}$,{ } EM output read noise.\\
%SNR$_{pc}$,{ } Signal to noise ratio in PC mode.\\
%SNR$_{lin}$,{ } SNR in linear mode.\\
%SNR$_{ideal}$,{ } SNR of a noise free detector.\\
%SNR$_{req}$,{ } required SNR in extracted spectrum.\\
%$\tau$,{ } required time resolution of observation.\\
%$T_{\text{SNR1}}$,{ }exposure time to reach SNR$_{lin}=1$ in extracted spectrum.\\
%$M$,{ } mean signal per pixel. \\
%$K$,{ } sky signal within one temporal bin. \\
%$D$,{ } dark charge within one temporal bin. \\
%$m$,{ } signal per-pixel-wavelength-step s$^{-1}$ . \\
%$t$,{ } photon-counting threshold (units e$_{pe}^-$).\\
%$n$,{ } mean counts in PC mode (counts pix$^{-1}$). \\
%$N$,{ } mean illumination (photo-electrons pix$^{-1}$). \\
%$d$,{ } spatial extent of spectrum (due to seeing), pixels.\\
%$\delta$,{ } photon-counting blocking factor.\\
%$\beta$,{ } off-chip binning factor.\\
%$\digamma$,{ } phase-folding factor.\\
%%%%%%%%%%%%%%%%%%%%%%%%%%%%%%%%%%%%%%%%
\section[]{Fractional charge of CICIR}
\label{sec:fractionalApp}
A CIC electron originating within the EM register will effectively have a fractional charge whose value is equal to the average charge $\bar q_O$ of a single photo-electron charge packet during its transit through the EM register.
The instantaneous charge $q_O$ of a pixel within the EM register that originated as a single photo-electron at the register input is:
\begin{equation}
q_O=(1+p)^x{  },
\label{eq:egrowth}
\end{equation}
where $x$ is the position within EM register and $p$ the per-transfer multiplication probability. The mean value, $\bar q_O$, of this pixel during its EM register transit
can then be obtained by integrating this function over the length of the register and
then dividing by the number of stages $S$. If we reference this charge to an equivalent signal at the
input to the register, such that $q_I=q_O/g_A$, we get:
\begin{equation}
\bar q_I=\dfrac{1}{Sg_A}\int_{x=1}^{S} (1+p)^x dx {  }.
\label{eq:hardmaths_a}
\end{equation}
This integral by the standard solution:
\begin{equation}
\bar q_I=\dfrac{1}{Sg_A\ln(1+p)}\left[(1+p)^x \right]_{x=1}^{S} {  },
\label{eq:hardmaths_b}
\end{equation}
which gives the result
\begin{equation}
\bar q_I=\frac{(1-g_A^{-1})}{\ln g_A}{  } \approx \dfrac{1}{\ln g_A} {  }.
\label{eq:hardmaths_c}
\end{equation}
Equation~\ref{eq:hardmaths_c} now allows us to calculate the per transfer probability of CIC $p_{C}$ in the EM register from a knowledge of the gain $g_A$ and the mean CIC charge in the bias $\nu_C$. The generation of this charge is a binomial process so $B_C$ the total number of CIC events in a pixel exiting the EM register is given by
\begin{equation}
B_C=p_{C}S{  }.
\label{eq:hardmaths_d}
\end{equation}
Since each of these CIC events will contain an average charge of $\bar q_I$, the mean per pixel CIC charge $\nu_C$ is given by
\begin{equation}
\nu_C=\bar q_I B_C {  },
\label{eq:hardmaths_e}
\end{equation}
substituting $\bar q_I$ from Equation~\ref{eq:hardmaths_d} and rearranging, we get
\begin{equation}
p_{C}\approx\frac{\ln(g_A)\nu_C}{S}{  }.
\label{eq:hardmaths_f}
\end{equation}
We can then show how the mean CIC charge in a pixel $\nu_C$ (relevant for linear-mode operation) and the mean number of CIC events in a pixel $B_C$ (relevant for PC operation) are related as follows:
\begin{equation}
\frac{B_C}{\nu_C}{  }\approx \ln(g_A){ }.
\label{eq:hardmaths_g}
\end{equation}
\section[]{SNR of an Ideal  Photon Counter}
\label{sec:PCnoiseDerivation}
We present a full derivation of SNR$_{pc}$, the SNR in a photon-counting detector. This differs from that of an ideal detector due to the effects of coincidence losses. To the best of our knowledge this derivation has not appeared before in the astronomical literature.

Let $N$ be the mean illumination in photo-electrons per pixel and $n$  the mean photon-counted signal per pixel, i.e. the fraction of pixels that contain one or more photo-electrons. Poisson statistics tells us that
\begin{equation}
n=1-e^{-N}{  },
\label{eq:neweq24}
\end{equation}
therefore:
\begin{equation}
N=-\ln(1-n){  }.
\label{eq:neweq25}
\end{equation}
The noise in a photon counted frame can be derived straightforwardly by considering that only two pixel values are possible: 0 and 1. Pixels containing 0 will have a variance of $N$, those containing 1 will have a variance of $N-1$. Knowing the fraction of pixels containing each of these two values then allows us to combine the variances in quadrature to yield $\sigma_{pc}$, the rms noise:
\begin{equation}
\sigma_{pc}=\sqrt{[e^{-N}N^2+(1-e^{-N})(N-1)^2]}{  },
\label{eq:neweq26}
\end{equation}
\begin{equation}
=\sqrt{(e^{-N}-e^{-2N})}{  }.
\label{eq:neweq27}
\end{equation}
The photon-counted images must then be processed to remove the effects of coincidence losses. This is done after the component frames within each temporal bin have been averaged to yield a mean value for $n$ for each pixel. The original mean signal $N$ prior to coincidence losses is then recovered by using Equation~\ref{eq:neweq25}. Although coincidence loss tends to produce a saturation and a smoothing of the image structure, the overall effect is to add a great deal of noise to the observation and for this reason we must avoid a photon-counting detector entering the coincidence loss regime. The amount of extra noise generated can be calculated by considering the change $dN$ in $N$ produced by a small change $dn$ in $n$. The noise in the final coincidence-corrected pixel will then be equal to that in the unprocessed average pixel multiplied by $dN/dn$. From Equation~\ref{eq:neweq25} we get
\begin{equation}
N+dN=-\ln[1-(n+dn)]{  }.
\label{eq:neweq29}
\end{equation}
This standard differential is then solved to yield
\begin{equation}
\frac{dN}{dn}=(1-n)^{-1}{  }.
\label{eq:neweq30}
\end{equation}
Substituting $N$ for $n$ using Equation~\ref{eq:neweq24} we get
\begin{equation}
\frac{dN}{dn}=e^{N}{  }.
\label{eq:neweq31}
\end{equation}
We then multiply the uncorrected  noise given in Equation~\ref{eq:neweq27} by this factor to yield the noise in the final coincidence-loss-corrected PC image. SNR$_{pc}$ is then given by:
\begin{equation}
\text{SNR}_{pc}=\frac{N}{\sqrt{e^N-1}}{  }.
\label{eq:neweq32}
\end{equation}
This can be expressed in units of $n$, using Equation~\ref{eq:neweq25}, which is more useful since it is $n$ that we actually measure from our images:
\begin{equation}
\text{SNR}_{pc}=\frac{-\ln(1-n)}{\sqrt{(1-n)^{-1}-1}}{  }.
\label{eq:neweq33}
\end{equation}
\section[]{SNR of a photon-counting EMCCD}
\label{sec:SNRderivationPC}
We have already shown  in Appendix \ref{sec:PCnoiseDerivation} that the SNR of an ideal photon counter is:
\begin{equation}
\text{SNR}_{pc}=\frac{N}{\sqrt{e^N-1}}{  },
\label{eq:psnrder1}
\end{equation}
with $N$ representing the signal per pixel.
This basic equation is now altered to a more realistic form to include the noise sources found in an EMCCD. Certain approximations are made during this process. The validity of these approximations have been verified by the Monte Carlo modelling.

So to begin with, we replace $N$ in the numerator with the detected fraction of photo-electrons at our chosen threshold and we replace $N$ in the denominator with the detected signal plus the detected CIC. In Figure~9 we have already shown that a threshold of 0.1e$^-_{pe}$ is close to optimum and that at this level 90\% of photo-electrons and 23\% of CICIR will be detected. We then get:
\begin{equation}
\text{SNR}_{pc}=\frac{0.9M}{\sqrt{\exp(0.9M+0.23B_C)-1}}{  },
\label{eq:psnrder2}
\end{equation}
where $M$ is the signal per temporal bin and $B_C$ the number of CICIR events per pixel.
We now need to consider that any photon counting observation will require the blocking of $\delta$ separate images if we are to achieve a usable SNR (the SNR of a single PC image with the exposure level lying within the sweet-spot is $\approx0.4$, see Figure~11). We then get:
\begin{equation}
\text{SNR}_{pc}=\frac{0.9M}{\sqrt{\delta}\sqrt{\exp(0.9M/\delta+0.23B_C)-1}}{  }.
\label{eq:psnrder3}
\end{equation}
Next we need to include other noise sources such as sky $K$ and dark charge $D$ (units of e$^-$ pix$^{-1}$).
Since these are indistinguishable from photo-electrons they will  have the same detected fraction for a given PC threshold. We also need to express the CICIR in terms of $\nu_C$ (see Equation~\ref{eq:hardmaths_g}), i.e. the mean per-pixel charge that CICIR contributes to the image. This is a parameter that can be measured directly  from the bias frames of an EMCCD camera. So we get:
\begin{equation}
\frac{0.9M}{\sqrt{\delta}\sqrt{\exp{[(0.9(M+D+K)/\delta)+0.23\ln(g_A)\nu_C]}-1}}.
\label{eq:SNRPC_EMCCD3}
\end{equation}
Note that $\nu_C$ the CICIR is multiplied by a factor of $\ln(g_A)$ in the denominator of Equation~\ref{eq:SNRPC_EMCCD3}. This is a consequence of the fractional charge of a CICIR electron (see Appendix~\ref{sec:fractionalApp}).

Note also that the read noise has been entirely ignored in this simplified description. This is fair since the threshold was set well above the noise, resulting in very few false counts. The read noise has a secondary influence, however, since it causes an effective blurring of the threshold level. This \textquoteleft fuzzy-threshold\textquoteright{ }must add some noise to the images since it can make all the difference as to whether an event lying close to the threshold is counted or not. The close fit between the approximation and the comprehensive model (see Figure~11) would, however, indicate that this noise source is not significant.
%This formula makes the important approximation that the detected fractions are fixed and independent of the %absolute signal levels. In the low signal regime this is a valid assumption but in regimes where there are %higher levels of signal and/or noise it will be less so. In the latter case, there is a significant %probability of a pixel containing more than a single electron; something that can only increase the %detected fraction.
\section[]{Model Variables}
\label{sec:MC_model}
The data cube describing the  count-rate from an EMCCD in PC-mode as a function of the mean illumination, the mean number of CIC events per pixel and the threshold setting, is available as an \texttt{IDL} .sav file, together with descriptions of the variables it contains, at this address: \protect{http://www.qucam.com/emccd/Histogram.html}\\ It is included for those wishing to check our results or to perform their own experiments.
Figure~D1 shows a selection of histograms extracted from the data cube by way of example.
\begin{figure}
\begin{center}
\includegraphics[width=0.45\textwidth]{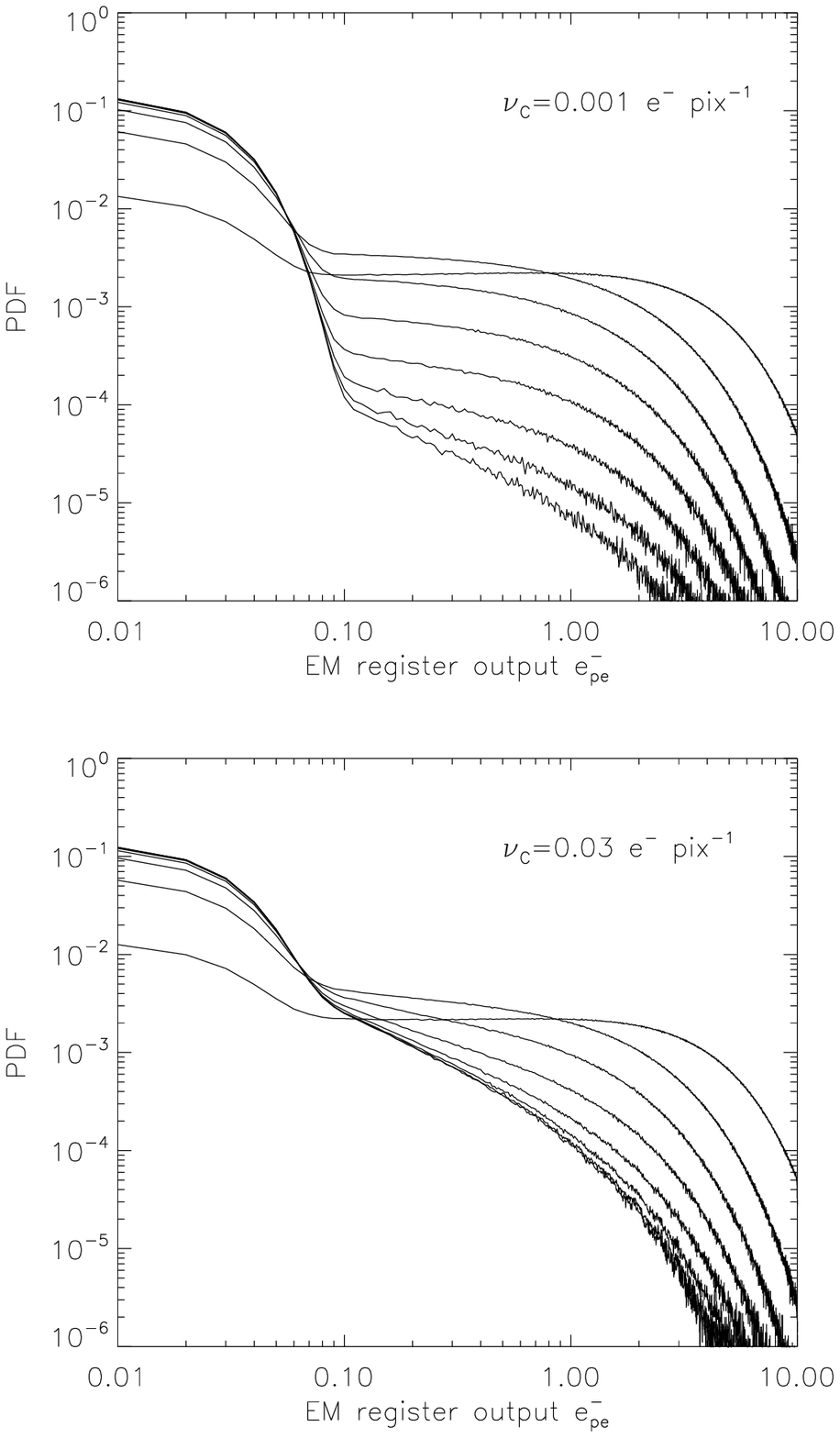}
\parbox{0.45\textwidth}{
\caption[MatrixP]{Some histograms generated by the EMCCD model, described in Section~4, are plotted here. These are taken from the \texttt{IDL} variable \texttt{outputMIXhist}. In the upper panel are a set of histograms with the signal level varying from 0.001 to 2.5e$^-$ pix$^{-1}$ and with the CIC fixed at $\nu_C=0.001$e$^-$ pix$^{-1}$. The lower plot has the same range of signals but with $\nu_C=0.03$e$^-$ pix$^{-1}$. The read noise was 0.025e$^-_{pe}$ in both cases.}}
\label{fig:probematrix}
\end{center}
\end{figure}
\clearpage
\section[]{Sky Backgrounds}
\label{sec:dark_app}
\begin{table}
\caption{WHT+ISIS sky backgrounds with QUCAM2.
Units: photo-electrons pix$^{-1}$s$^{-1}$.   Slit width= 1". From the ING web pages.}
\begin{center}
    \begin{tabular}{|l|l|l }
    \hline
    Grating & Bright & Dark  \\
    \hline
    R316R & 0.12 & 0.014  \\
    R600R & 0.05 & 0.006  \\
    R1200R & 0.023 & 0.003 \\
    R300B & 0.07 & 0.004  \\
    R600B & 0.03 & 0.002  \\
    R1200B & 0.016 & 0.001 \\
    \hline
    \end{tabular}
\end{center}
\label{tab:dark_tableWHT}
\end{table}
\begin{table}
\caption{Estimated sky backgrounds for an E-ELT+ISIS-type instrument with VPH gratings. The plate scale and the detector QE is assumed to be the same as for QUCAM2 on ISIS.
Units: photo-electrons pix$^{-1}$s$^{-1}$.   Slit width= 1".}
\begin{center}
    \begin{tabular}{|l|l|l }
    \hline
    Grating & Bright & Dark  \\
    \hline
    Red 316 & 15 & 1.6  \\
    Red 600 & 6.0 & 0.7  \\
    Red 1200 & 2.7 & 0.4 \\
    Blue 300 & 8.2 & 0.47  \\
    Blue 600 & 3.6 & 0.23  \\
    Blue 1200 & 1.9 & 0.12 \\
    \hline
    \end{tabular}
\end{center}
\label{tab:dark_tableELT}
\end{table}
%%%%%%%%%%%%%%%%%%%%%%%%%%%%%%%%%%%%%%%%
\bsp
\label{lastpage}
\end{document}